\documentstyle[12pt,cite,epsf]{article}
\def\Thebibliography#1{\section*{References}
 \list
 {[\arabic{enumi}]}{
\addtolength{\baselineskip}{-0.7\baselineskip}
\setlength{\parsep}{0.0ex}
\settowidth\labelwidth{[#1]}\leftmargin\labelwidth
 \advance\leftmargin\labelsep
 \usecounter{enumi}}
 \def\newblock{\hskip .11em plus .33em minus .07em}
 \sloppy\clubpenalty4000\widowpenalty4000
 \sfcode`\.=1000\relax}

\let\Large=\large
\let\large=\normalsize
\newcommand{\eqn}[1]{(\ref{#1})}

\newcommand{\be}{\begin{equation}}
\newcommand{\ee}{\end{equation}}
\newcommand{\ba}{\begin{array}{c}}
\newcommand{\bat}{\begin{array}{cc}}
\newcommand{\ea}{\end{array}}
\newcommand{\beqn}{\begin{eqnarray}}
\newcommand{\eeqn}{\end{eqnarray}}

\newcommand{\bi}{\begin{itemize}}
\newcommand{\ei}{\end{itemize}}

\newcommand{\rms}{\rm\scriptsize}

\newcommand{\ssb}{\stackrel{\mbox{\rms SSB}}{\longrightarrow}}
\newcommand{\toU}{\stackrel{\mbox{\rms U(1)}}{\longrightarrow}}
\newcommand{\toG}{\stackrel{\mbox{\rms G}}{\longrightarrow}}
\newcommand{\toLow}{\stackrel{q^2<<M_W^2}{\,\longrightarrow\,}}
\newcommand{\vecW}{\stackrel{\rightarrow}{W}}
\newcommand{\vecWp}{\stackrel{\rightarrow}{W'}}
\newcommand{\toTheta}{\stackrel{\vec{\theta}=\vec{0}}{\longrightarrow}}
\newcommand{\vecWmn}{\stackrel{\!\!\!\!\!\!\rightarrow}{W^{\mu\nu}}}
\newcommand{\simbigs}{\stackrel{s\to\infty}{\,\sim\,}}
\newcommand{\cL}{{\cal L}}

\newcommand{\cM}{{\cal M}}
\newcommand{\cO}{{\cal O}}
\newcommand{\cP}{{\cal P}}

\newcommand{\cH}{{\cal H}}
\newcommand{\cA}{{\cal A}}
\newcommand{\cI}{{\cal I}}
\newcommand{\cF}{{\cal F}}
\newcommand{\bV}{\mbox{\boldmath $V$}}

\newcommand{\no}{\nonumber}
\newcommand{\bel}[1]{\be\label{#1}}
\newcommand{\e}{\mbox{\rm e}}

\begin{document}

\begin{titlepage}
\begin{flushright}
FTUV/94-62\\ IFIC/94-59\\  hep-ph/9412274 \\                    
\end{flushright}
\vspace*{1cm}
\begin{center}
{\Large \bf  THE STANDARD MODEL OF \\ ELECTROWEAK INTERACTIONS}
\\
\vspace*{0.5cm}
{\bf A. Pich}
\\
Departament de F\'\i sica Te\`orica
and IFIC,  Universitat de Val\`encia -- CSIC\\
Dr. Moliner 50,
E--46100 Burjassot, Val\`encia, Spain \\
\vspace*{4.5cm}
{\Large\bf   Abstract  \\ }
\end{center}

What follows is an updated version of the lectures given at the
CERN Academic Training (November 1993) and at the Jaca Winter
Meeting  (February 1994).
The aim is to provide a pedagogical introduction to the
Standard Model of electroweak interactions.
After briefly reviewing the empirical considerations which lead to the
construction of the Standard Model Lagrangian, the particle content,
structure and symmetries of the theory are discussed.
Special emphasis is given to the many phenomenological tests
(universality, flavour-changing neutral currents, precision
measurements, quark mixing, etc.)
which have established this theoretical framework
as the Standard Theory of electroweak interactions.

\vspace*{1.7cm}
\begin{center}
Lectures given at the 
XXII International Winter Meeting on Fundamental Physics,\\
{\em The Standard Model and Beyond}, 
Jaca (Spain), 7-11 February 1994, \\
and at the \\
CERN Academic Training,
 Geneva (Switzerland),
15-26 November 1993
\end{center}

\vfill
\begin{flushleft}
FTUV/94-62\\
IFIC/94-59\\
November 1994
\end{flushleft}
\end{titlepage}



\section{Introduction}

The Standard Model (SM)
is a gauge theory, based on the group
%
$SU(3)_C \otimes SU(2)_L \otimes U(1)_Y$,
%
which describes strong, weak and electromagnetic interactions,
via the exchange of the corresponding spin--1 gauge fields:
8 massless gluons and 1 massless photon for the strong and
electromagnetic interactions, respectively,
and 3 massive bosons, $W^\pm$ and $Z$, for the weak interaction.
The fermionic-matter content is given by the known
leptons and quarks, which are organized in a 3--fold
family structure:
\bel{eq:families}
\left[\bat \nu_e & u \\  e^- & d \ea \right] \, , \qquad\quad
\left[\bat \nu_\mu & c \\  \mu^- & s \ea \right] \, , \qquad\quad
\left[\bat \nu_\tau & t \\  \tau^- & b \ea \right] \, ,
\ee
where
(each quark appears in 3 different ``colours'')
\bel{eq:structure}
\left[\bat \nu_l & q_u \\  l^- & q_d \ea \right] \,\,\equiv\,\,
\left(\ba \nu_l \\ l^- \ea \right)_L , \quad
\left(\ba q_u \\ q_d \ea \right)_L , \quad l^-_R ,
\quad (q_u)_R , \quad
(q_d)_R ,
\ee
plus the corresponding antiparticles.
Thus, the left-handed fields are $SU(2)_L$ doublets, while
their right-handed partners transform as $SU(2)_L$ singlets.
The 3 fermionic families in Eq.~\eqn{eq:families} appear
to have identical properties (gauge interactions); they only
differ by their mass and their flavour quantum number.

The gauge symmetry is broken by the vacuum,
which triggers the Spontaneous Symmetry Breaking (SSB)
of the electroweak group to the electromagnetic subgroup:
\bel{eq:ssb}
SU(3)_C \otimes SU(2)_L \otimes U(1)_Y \, \ssb\,
SU(3)_C \otimes U(1)_{QED} \, .
\ee
The SSB mechanism generates the masses of the weak gauge bosons,
and gives rise to the appearance of
a physical scalar particle in the model, the
so-called ``Higgs''.

The SM constitutes one of the most successful achievements
in modern physics. It provides a very elegant theoretical
framework, which is able to describe {\em all} known experimental
facts in particle physics.

These lectures provide an introduction to the
electroweak sector of the SM, i.e. the
$SU(2)_L \otimes U(1)_Y$ part
\cite{GL:61,WE:67,SA:69,GIM:70}
(the strong $SU(3)_C$ piece is discussed in
Ref.~\cite{sorrento}).
Sects. \ref{sec:history}
and \ref{sec:problems} describe
some experimental and theoretical
arguments suggesting the structure presented above
[Eqs.~\eqn{eq:families} to \eqn{eq:ssb}]
as the natural model for describing the electroweak interactions.
The power of the gauge principle is shown in
Sect.~\ref{sec:qed}, in
the simpler QED case.
The SM framework is presented in Sects. \ref{sec:model},
\ref{sec:ssb} and \ref{sec:flavour},
which discuss the gauge structure, the SSB mechanism and
the family structure, respectively.
Some further theoretical considerations concerning quantum
anomalies are given in Sect.~\ref{sec:anomalies}.
Sects. \ref{sec:tree} to \ref{sec:cc-leptons} summarize
the present phenomenological status of the SM.
A few comments  on open questions, to be tested at future facilities,
are finally given in Sect.~\ref{sec:summary}.


\setcounter{equation}{0}
\section{Low-Energy Experimental Facts}
\label{sec:history}

\subsection{$\mu^-\to e^-\bar\nu_e\nu_\mu$ decay}
\label{subsec:mu-decay}

Let us parametrize the  3--body decay of the muon by a general
local, derivative-free, 4--fermion Hamiltonian:
\bel{eq:mu_hamiltonian}
\cH_{\mbox{\rms eff}}\, = \,
\sum_{n,\epsilon,\omega}\, g^n_{\epsilon,\omega}\,
\left[\bar e_\epsilon\Gamma^n (\nu_e)_\sigma\right]\,
\left[(\bar\nu_\mu)_\lambda\Gamma_n \mu_\omega\right]\, .
\ee
Here, $\epsilon,\omega,\sigma,\lambda$ denote the chiralities
(left-handed, right-handed) of the corresponding fermions,
and $n$ labels the type on interaction: scalar ($I$),
vector ($\gamma^\mu$), tensor($\sigma^{\mu\nu}$).
For given $n,\epsilon,\omega$, the neutrino chiralities $\sigma$
and $\lambda$ are uniquely determined.

The couplings $g^n_{\epsilon,\omega}$ can be determined
experimentally, by studying the energy and angular
(with respect to the $\mu^-$-spin) distribution of the
final electron, the $e^-$ polarization,
and the cross-section of the related
$\nu_\mu e^-\to\mu^-\nu_e$ process.
One finds\footnote{
The most recent analysis \protect\cite{fgj:86}
finds that the probability of having a left-handed $\mu^-$
decaying into a left-handed $e^-$ is bigger than
95\% (90\% CL).}
that the decay amplitude involves only  left-handed
fermions, with an effective Hamiltonian of the $V-A$ type:
\bel{eq:mu_v_a}
\cH_{\mbox{\rms eff}}\, = \, {G_F \over\sqrt{2}}
\left[\bar e\gamma^\alpha (1-\gamma_5) \nu_e\right]\,
\left[ \bar\nu_\mu\gamma_\alpha (1-\gamma_5)\mu\right]\, .
\ee

The so-called Fermi coupling constant $G_F$
is fixed by the total decay width,
\bel{eq:mu_lifetime}
{1\over\tau_\mu}\, = \, \Gamma(\mu^-\to e^-\bar\nu_e\nu_\mu)
\, = \, {G_F^2 m_\mu^5\over 192 \pi^3}\,
\left( 1 + \delta_{\mbox{\rms QED}}\right) \,
f\left({m_e^2\over m_\mu^2}\right) \, ,
\ee
where
$\, f(x) = 1-8x+8x^3-x^4-12x^2\ln{x}$,
and
%
$\,\delta_{\mbox{\rms QED}}  =  {\alpha\over 2\pi}\left({25\over 4}
-\pi^2\right)  \approx  -0.0042 \, $
%
takes into account the leading radiative QED corrections
\cite{KS:59}.
{}From the measured lifetime \cite{pdg:94},
$\tau_\mu=(2.19703\pm 0.00004)\times 10^{-6}$ s,
one gets the value
\bel{eq:gf}
G_F\, = \, (1.16639\pm 0.00002)\times 10^{-5} \,\mbox{\rm GeV}^{-2}
\,\approx\, {1\over (293 \,\mbox{\rm GeV})^2} \, .
\ee

\subsection{Beta decay}
\label{subsec:beta-decay}

The weak transitions $d\to u e^-\bar\nu_e$ and $u\to d e^+\nu_e$
can be studied through the corresponding hadronic decays
$n\to p e^-\bar\nu_e$ and $p\to n e^+\nu_e$, where the last
process can only occur within a nuclear transition because
it is kinematically forbidden for a free proton.
The experimental analysis of these processes shows that they
can be described by the effective Hamiltonian
\bel{eq:beta_decay}
\cH_{\mbox{\rms eff}}\, = \, {G^{\Delta S=0} \over\sqrt{2}}
\left[\bar p\gamma^\alpha (1-g_A\gamma_5) n\right]\,
\left[ \bar e\gamma_\alpha (1-\gamma_5)\nu_e\right]\, ,
\ee
where \cite{pdg:94}
\bel{eq:G}
G^{\Delta S=0}\,\approx \, 0.975\, G_F\, ,
\qquad\qquad
g_A\, = \, 1.2573\pm0.0028\, .
\ee
The strength of the interaction turns out to be approximately
the same as for $\mu$ decay and, again,
only left-handed leptons are involved.
The strong similarity with
Eq.~\eqn{eq:mu_v_a} suggest a universal
(same type and strength) interaction at the
quark-lepton level:
\bel{eq:quark}
\cH_{\mbox{\rms eff}}\, = \, {G^{\Delta S=0} \over\sqrt{2}}
\left[\bar u\gamma^\alpha (1-\gamma_5) d\right]\,
\left[ \bar e\gamma_\alpha (1-\gamma_5)\nu_e\right]\, .
\ee
In fact, the conservation of the vector current,
$\partial_\mu\left(\bar u\gamma^\mu d\right) = 0$,
implies
%
$\,\langle p|\bar u\gamma^\mu d|n\rangle 
 =  \bar p \gamma^\mu n \, $
%
at $q^2=0$;
i.e. strong interactions do not renormalize\footnote{
This is completely analogous to the
electromagnetic-charge conservation in QED:
the conservation of the
electromagnetic current implies that the proton electromagnetic
form factor does not get any QED or QCD correction at $q^2=0$, and,
therefore,
$Q(p)=2Q(u)+Q(d)=|Q(e)|$.}
the vector current.
However, the axial-current matrix elements do get modified
by the QCD dynamics.
Thus, the factor $g_A$ can be easily understood\footnote{
The conservation of the vector and axial currents is associated
with the chiral symmetry of the QCD Lagrangian
\protect\cite{sorrento}. Chirality is however not respected
by the QCD vacuum.
The SSB of the axial generators gives rise to massless Goldstone
bosons
(see Sect.~\protect\ref{subsec:goldstone}), the pions,
which couple to the axial currents.
One can easily derive the approximate (Goldberger--Treiman)
relation:
$g_A\approx g_{\pi NN} f_\pi/M_N \approx 1.3$,
where $g_{\pi NN}$ is the strength of the $\pi NN$ interaction
and $f_\pi$ ($=92.4$ MeV) the pion decay constant.
}
as a QCD effect.
The interaction \eqn{eq:quark}
correctly describes the weak decay
$\pi^+\to\pi^0e^+\nu_e$ (Br = $(1.025\pm 0.034)\times 10^{-8}$
\cite{pdg:94}).

\subsection{$\pi^-\to l^-\bar\nu_l$}
\label{subsec:pi_decay}

One finds experimentally that the final charged lepton in the
2--body $\pi^-$ decay is always right-handed. By angular-momentum
conservation,
the $\bar\nu_l$ is also right-handed.
If one assumes that only left-handed leptons (and right-handed
anti-leptons) participate in the weak interaction, the
$\pi^-\to l^-\bar\nu_l$ decay
should be forbidden in the limit of zero lepton massess
(helicity is then a good quantum number).
The interaction \eqn{eq:quark} predicts in fact a strong
helicity suppression of these decays \cite{MS:93},
\bel{eq:r_e_mu}
R_{e/\mu}\,\equiv\, {\Gamma(\pi^-\to e^-\bar\nu_e)\over
\Gamma(\pi^-\to \mu^-\bar\nu_\mu)}\, = \,
{m_e^2 (1-m_e^2/m_\pi^2)^2\over m_\mu^2 (1-m_\mu^2/m_\pi^2)^2}
\, (1+\delta_{\mbox{\rms QED}})
\, = \, (1.2352\pm 0.0005)\times 10^{-4} \, ,
\ee
in excellent agreement with the
measured ratio
$R_{e/\mu} = (1.230\pm 0.004)\times 10^{-4}$
\cite{pdg:94}.

\subsection{Neutrino flavours}
\label{subsec:neutrinos}

If the two neutrinos produced in the
$\mu^-\to e^-\nu_\mu\bar\nu_e$ decay
had the same lepton flavour, i.e.
$\nu_e=\nu_\mu$,
one could
contract the two neutrino legs in Eq.~\eqn{eq:mu_v_a}
and generate (provided one is able to make sense of the divergent
neutrino loop!)
a $\mu^-\to e^-\gamma$ transition, by simply radiating
a photon from the charged-lepton lines.
The strong experimental upper-limit on this decay \cite{pdg:94},
Br$(\mu^-\to e^-\gamma)< 4.9\times 10^{-11}$ (90\% CL),
provides then significant evidence of the existence of
different neutrino flavours.

A direct experimental test can be obtained with neutrino beams.
The decay $\pi^-\to \mu^-\bar\nu_\mu$ can be used to
produce a $\bar\nu_\mu$ beam, out of a parent beam of pions.
Studying the interactions of this neutrino beam with matter,
one observes \cite{DA:62}
that only $\mu^+$ are produced, but not $e^+$:
\bel{eq:nu_X}
\bar\nu_\mu X \to \mu^+ X' \, ,
\qquad\qquad
\bar\nu_\mu X \not\to e^+ X' \, .
\ee
Analogously, a beam of $\bar\nu_e$ produces $e^+$ but never
$\mu^+$.
Therefore, the neutrino partners of the electron and the muon
are two different particles:
%
$\nu_e\not=\nu_\mu $.
%

\subsection{$\Delta S=1$ transitions}
\label{subsec:kaons}

The analysis of strangeness-changing decays
[$K\to (\pi) l^-\bar\nu_l$,
$\Lambda\to p e^-\bar\nu_e$, \ldots] shows that:
\bi
\item The weak interaction is always of the $V-A$ type.
\item The strength of the interaction is the same in all
decays; however, it is smaller than the one measured in
$\Delta S=0$ processes:
\bel{eq:cabibbo}
G^{\Delta S=1} \, \approx \, 0.22 \,  G_F \, .
\ee
\item All decays satisfy the $\Delta S=\Delta Q$ rule
[i.e. decays such as $\Sigma^+\to n e^+\nu_e$
or $\bar K^0\to\pi^-l^+\nu_l$ never occur],
as expected from a $s\to u l^-\bar\nu_l$ transition.
\ei

\subsection{The $V-A$ model}
\label{subsec:v-a}

All previous experimental facts can be nicely described by the
Hamiltonian:
\bel{eq:v_a}
\cH \, = \, {G_F\over\sqrt{2}}\, J^\mu J_\mu \, ,
\ee
where
\bel{eq:current}
J^\mu\, =\,
\bar u \gamma^\mu (1-\gamma_5)
\left[\cos{\theta_C} d + \sin{\theta_C} s\right]
+ \bar\nu_e \gamma^\mu (1-\gamma_5) e
+ \bar\nu_\mu \gamma^\mu (1-\gamma_5) \mu \, .
\ee
Thus, at low-energies, weak transitions proceed through
a universal interaction (the same for all fermions),
involving charged-currents only. The different strength of hadronic
$\Delta S=0$ and $\Delta S=1$ processes can be simply
understood \cite{cabibbo}
as originating from the mixing angle $\theta_C$,
defined as
$\sin{\theta_C}\equiv G^{\Delta S=1} /G_F \approx 0.22$. Thus,
the weak partner of the up-quark is a mixture of $d$ and $s$.
Note, that $\cos{\theta_C}\approx 0.975$ in agreement with
Eq.~\eqn{eq:G}.

\setcounter{equation}{0}
\section{High-Energy Behaviour}
\label{sec:problems}

At high energies, the Hamiltonian \eqn{eq:v_a}
cannot be a correct description of weak interactions.
There are two fundamental problems with the
$V-A$ interaction:
\begin{enumerate}
\item {\bf Renormalizability:}
Higher-order (loop) transitions such as
$\nu_\mu e^-\to\mu^-\bar\nu_e\to\nu_\mu e^-$
are divergent
[$T\sim\int\,d^4k\, (1/k^2) = \infty$].
Ultraviolet loop divergences also occur in well-behaved
Quantum Field Theories like QED; but, there, all infinities
can be eliminated through a redefinition of parameters
(renormalization),
so that measurable quantities are always finite.
The problem with the interaction \eqn{eq:v_a} is that it is
non-renormalizable: it is impossible to eliminate all
infinities by simply redefining the parameters and fields.
\item {\bf Unitarity:}
Even at tree-level, the $V-A$ Hamiltonian predicts a bad
high-energy behaviour. Since $G_F$ is a dimensionful
quantity
($[G_F] = M^{-2}$),
the interaction \eqn{eq:v_a} gives rise to cross-sections
which increase with energy:
\bel{eq:he_cs}
\sigma(\nu_\mu e^-\to\mu^-\nu_e)\,\approx\,
G_F^2 s/\pi \, .
\ee
At large values of $s$, unitarity is clearly violated
(the probability of the transition is bigger than 1).
The unitarity bound
$\sigma<2\pi/s$ is only satisfied if
$s\leq \sqrt{2}\pi/G_F\sim (617\,\mbox{\rm GeV})^2$.
\end{enumerate}
Therefore, the succesful $V-A$ model can only be a low-energy
effective theory of some more fundamental dynamics.

\subsection{Intermediate Vector Boson hypothesis}
\label{subsec:ivb}

\begin{figure}[tbh]
\vfill
\centerline{
\begin{minipage}[t]{.47\linewidth}
\centerline{\mbox{\epsfysize=4.0cm\epsffile{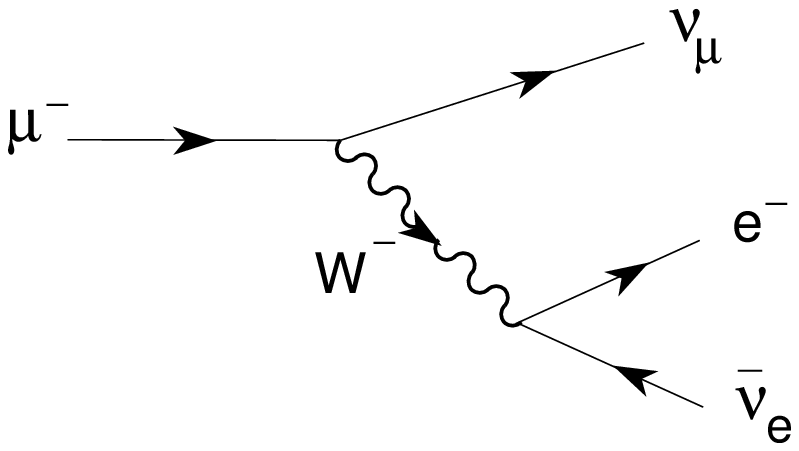}}}
\caption{$\mu$-decay diagram.}
\label{fig:mudecay}
\end{minipage}
\hspace{1.0cm}
\begin{minipage}[t]{.47\linewidth}
\centerline{\mbox{\epsfysize=4.0cm\epsffile{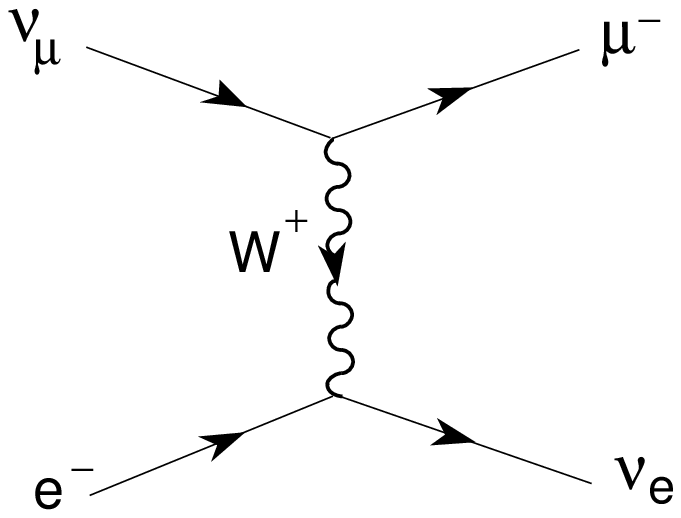}}}
\caption{$\nu_\mu e^-\to\mu^-\nu_e$ through $W$ exchange.}
\end{minipage}
}
\vfill
\end{figure}

In QED the fundamental $\gamma \bar e e$ interaction
generates 4-fermion couplings through $\gamma$-exchange.
However, since the photon is massless, the resulting
$\bar ee \bar ee$ interaction is non local; the photon
propagator gives rise to a long-range force, with an amplitude
$T\sim\alpha/q^2$.
Since weak interactions are short-range, we would rather need
some massive object to play the role of the photon in QED.
If one assumes \cite{LY:60}
that the charged current couples to a massive
spin-1 field $W_\mu$,
\bel{eq:ivh}
\cL\, =\, {g\over 2\sqrt{2}}\, \left( J^\mu W^\dagger_\mu +
\mbox{\rm h.c.}\right) \, ,
\ee
the $V-A$ interaction can be generated through
$W$-exchange.
At energies much lower than the $W$ mass, the vector-boson
propagator reduces to a contact interaction,
\bel{eq:low_energy}
{-g_{\mu\nu} + q_\mu q_\nu/M_W^2 \over q^2-M_W^2}\quad
 \toLow\quad {g_{\mu\nu}\over M_W^2}\, .
\ee
Eq.~\eqn{eq:v_a} is then obtained with the identification
\bel{eq:g_G}
{g^2\over 8 M_W^2}\, = \, {G_F\over\sqrt{2}}\, .
\ee
In order to have a perturbative coupling, i.e. $g<1$,
the massive intermediate boson should satisfy
$M_W<123$ GeV.

The interaction \eqn{eq:ivh} gives rise to a better
high-energy behaviour for $\nu l^-\to \nu l^-$, 
\bel{eq:he_cs_ivb}
\sigma(\nu_\mu e^-\to\mu^-\nu_e) \;\;\simbigs\;\;
G_F^2 M_W^2/\pi \, .
\ee
Although there is still a violation of unitarity, the
cross-section does not grow any longer with energy.
However, the unphysical rise of the cross-section
reappears now in those processes where longitudinal
$W$ bosons are produced:
\bel{eq:w_production}
\sigma(\nu_e\bar\nu_e\to W^+_L W^-_L)\;\;\simbigs\;\;  s\, ;
\qquad\qquad
\sigma(e^+e^-\to W^+_L W^-_L)\;\;\simbigs\;\;  s\, .
\ee

The origin of the problem can be better understood analyzing the
1--loop box amplitude $T(e^+e^-\to W^+ W^-\to e^+e^-)$, where
the $W$ fields appear virtually
[the absorptive part of this amplitude is related to the $e^+e^-\to
W^+ W^-$ production process].
The bad high-energy behaviour stems from the
$q_\mu q_\nu$ piece of the $W$ propagators, which gives rise to a
quadratically divergent loop integral
[$T\sim\int\,d^4q /q^2=\infty$].
A similar diagram exists in QED, with photons instead of $W$'s;
however, the conservation of the electromagnetic current,
$q_\mu J^\mu_{\mbox{\rms em}}=0$, makes those $q^\mu$ contributions
harmless. The absence of this problem in QED is related to
the associated gauge symmetry [see Sect.~\ref{sec:qed}],
which requires a massless photon.

A possible way out would be the existence of an additional
contribution to the $W^+ W^-$ production amplitudes, which cancels
the rising of the cross-section at large energies.
In fact, since the $W$'s have electric charge, one should also consider
the s-channel contribution $e^+e^-\to\gamma\to W^+ W^-$,
which gives rise to a similar $\sigma\sim s$ behaviour.
The bad  high-energy behaviour
could be eliminated from the sum of the
weak and electromagnetic amplitudes, provided that the
weak coupling $g$
and the electromagnetic coupling $e$ are related;
this points towards some kind of electroweak unification.
However, even if one succeeds to realize this cancellation,
the problem still remains in the
$\nu_e\bar\nu_e\to W^+ W^-$ production amplitude, because the
photon does not couple to neutrinos.

\subsection{Neutral currents}
\label{subsec:nc}

The high-energy cancellation can be realized introducing an
additional
neutral intermediate boson $Z$, which couples both to neutrinos and
charged leptons.
By cleverly choosing the $Z$ mass and couplings, it is possible
to obtain a cancellation with
the s-channel contributions
$e^+e^-\to Z\to W^+ W^-$ and
$\nu_e\bar\nu_e\to Z\to W^+W^-$.
This idea has important implications.

The exchange of a $Z$ boson in the t channel, should give rise
to neutral-current processes such as
$\nu_\mu e^-\to \nu_\mu e^-$ or $\nu_\mu p\to \nu_\mu p$.
The experimental confirmation of this kind of phenomena
was done in 1973 \cite{HA:73}, with the discovery
of the elastic scattering
\bel{eq:nc_scattering}
\nu_\mu e^-\,\to\,\nu_\mu e^-.
\ee

Neutral-current interactions
have been extensively analyzed in many experiments.
In contrast with the charged-current transitions, one finds that
flavour-changing neutral-current processes are very suppressed
\cite{pdg:94}:
\beqn\label{eq:fcnc}
{\Gamma(\mu^-\to e^-e^+e^-)\over \Gamma(\mu^-\to e^-\bar\nu_e\nu_\mu)}
< 1.0\times 10^{-12};
\qquad & &
{\Gamma(\Sigma^+\to p e^+e^-)\over \Gamma(\Sigma^-\to n e^-\bar\nu_e)}
< 1.3\times 10^{-2};
\qquad (90\% \mbox{\rm CL}) .
\no\\
{\Gamma(K_L\to \mu^+\mu^-)\over \Gamma(K^+\to \mu^+\nu_\mu)}
& =& (2.79\pm 0.15)\times10^{-9} .
\eeqn
%
Therefore, the  $Z$ couplings 
are flavour diagonal.

\subsection{Wanted ingredients for a theory of weak interactions}
\label{subsec:ingredients}

The  previous experimental and theoretical arguments,
suggest a series of requirements that the fundamental
theory of weak interactions should satisfy:
\bi
\item Intermediate spin--1
bosons $W^\pm$, $Z$ and $\gamma$.
\item Electroweak unification:
$g_W/2\sqrt{2}\sim g_Z/2\sqrt{2}\sim e$,
i.e. $g^2/4\pi\sim 8\alpha$.
Together with the relation \eqn{eq:g_G},
the unification of couplings implies
\bel{eq:mW}
M_W\sim\biggl({\sqrt{2} g^2\over 8 G_F}\biggr)^{1/2}
\sim
\biggl({4\pi\alpha\sqrt{2}\over G_F}\biggr)^{1/2}\sim
100\, \mbox{\rm GeV} .
\ee
\item The $W^\pm$ field couples only to left-handed doublets
\bel{eq:W-doublets}
\left(\ba \nu_e \\ e^- \ea \right)_L , \qquad
\left(\ba \nu_\mu \\ \mu^- \ea \right)_L , \qquad
\left(\ba u \\ d_C \ea \right)_L ,
\ee
where
$d_C\equiv\cos{\theta_C} d + \sin{\theta_C} s$.
\item The $Z$ boson has only flavour-diagonal couplings.
\item Lepton-number is conserved.
\item Renormalizability.
In order to satisfy this requirement, we need a gauge theory.
\ei

\setcounter{equation}{0}
\section{Gauge Symmetry: QED}
\label{sec:qed}

Let us consider the Lagrangian describing a free Dirac fermion:
\bel{eq:l_free}
\cL_0\, =\, i \,\overline{\Psi}(x)\gamma^\mu\partial_\mu\Psi(x)
\, - \, m\, \overline{\Psi}(x)\Psi(x) \, .
\ee
$\cL_0$ is invariant under {\em global}
$U(1)$ transformations
\bel{eq:global}
\Psi(x) \,\toU\, \Psi'(x)\,\equiv\,\exp{\{i Q \theta\}}\,\Psi(x) \, ,
\ee
where $Q\theta$ is an arbitrary real constant.
Clearly, the phase of $\Psi(x)$ is a pure convention-dependent
quantity without physical meaning.

However,
the free Lagrangian is no-longer invariant if one allows
the phase transformation to depend on the space-time coordinate,
i.e. under {\em local} phase redefinitions $\theta=\theta(x)$,
because
\bel{eq:local}
\partial_\mu\Psi(x) \,\toU\, \exp{\{i Q \theta\}}\,
\left(\partial_\mu + i Q \partial_\mu\theta\right)\,
\Psi(x) \, .
\ee
Thus, once an observer situated at the point $x_0$
has adopted a given phase-convention, the same convention must
be taken at all space-time points. This looks very unnatural.

The ``Gauge Principle'' is the requirement that the $U(1)$
phase invariance should hold {\em locally}.
This is only possible if one adds some additional piece to the
Lagrangian, transforming in such a way  as to cancel the
$\partial_\mu\theta$ term in Eq.~\eqn{eq:local}.
The needed modification is completely fixed by the transformation
\eqn{eq:local}: one introduces a new spin--1
(since $\partial_\mu\theta$  has a Lorentz index)
field $A_\mu(x)$, transforming as
\bel{eq:a_transf}
A_\mu(x)\,\toU\, A_\mu'(x)\,\equiv\, A_\mu(x) + {1\over e}\,
\partial_\mu\theta\, ,
\ee
and defines the covariant derivative
\bel{eq:d_covariant}
D_\mu\Psi(x)\,\equiv\,\left[\partial_\mu-ieQA_\mu(x)\right]
\,\Psi(x)\, ,
\ee
which has the required 
property of transforming like the field itself:
\bel{eq:d_transf}
D_\mu\Psi(x)\,\toU\,\left(D_\mu\Psi\right)'(x)\,\equiv\,
\exp{\{i Q \theta\}}\,D_\mu\Psi(x)\,.
\ee
The Lagrangian
\bel{eq:l_new}
\cL\,\equiv\,
i \,\overline{\Psi}(x)\gamma^\mu D_\mu\Psi(x)
\, - \, m\, \overline{\Psi}(x)\Psi(x)
\, =\, \cL_0\, +\, e Q A_\mu(x)\, \overline{\Psi}(x)\gamma^\mu\Psi(x)
\ee
is then invariant under local $U(1)$ transformations.

The gauge principle has automatically generated an interaction
term between the Dirac spinor and the gauge field $A_\mu$,
which is nothing else than the familiar QED interaction.
Note that the corresponding electromagnetic
charge $eQ$ is completely arbitrary.
If one wants $A_\mu$ to be a true propagating field, one needs to add
a gauge-invariant kinetic term
\bel{eq:l_kinetic}
\cL_{\mbox{\rms Kin}}\,\equiv\, -{1\over 4} F_{\mu\nu} F^{\mu\nu}\,,
\ee
where
$F_{\mu\nu}\,\equiv\, \partial_\mu A_\nu -\partial_\nu A_\mu$
is the usual electromagnetic field strength.
A possible mass term for the gauge field,
${1\over 2}m^2A^\mu A_\mu$, is forbidden because it would violate
gauge invariance; therefore,
the photon field is predicted to be massless.

The total Lagrangian in Eqs.~\eqn{eq:l_new} and \eqn{eq:l_kinetic}
gives rise to the well-known Maxwell equations.
{}From our gauge symmetry requirement, we have deduced
the right QED Lagrangian, which leads to a very  successful
quantum field theory.
Remember that QED predictions have been tested
to a very high accuracy, as exemplified by the electron and
muon anomalous magnetic moments
[$a_l\equiv (g_l-2)/2$, where $\mu_l\equiv g_l \,(e \hbar/2m_l)$]
\cite{KI:90}:
\beqn\label{eq:a_e}
a_e&=&\left\{ \bat
(115 \, 965 \, 214.0\pm 2.8) \times 10^{-11} & (\mbox{\rm Theory})
\\
(115 \, 965 \, 219.3\pm 1.0) \times 10^{-11} & (\mbox{\rm Experiment})
\ea \, , \right.\\
a_\mu&=&\left\{ \bat
(1 \, 165 \, 919.2\pm 1.9)
\times 10^{-9} & (\mbox{\rm Theory})
\\
(1 \, 165 \, 923.0\pm 8.4) \times 10^{-9} & (\mbox{\rm Experiment})
\ea \, . \right.
\eeqn
%


\setcounter{equation}{0}
\section{The $SU(2)_L\otimes U(1)_Y$ Theory}
\label{sec:model}

To describe weak interactions, we need a more elaborated
structure, with several fermionic flavours and
different properties for left- and right-handed fields.
Moreover, the left-handed fermions should appear in doublets,
and we would like to have massive gauge bosons $W^\pm$ and
$Z$ in addition to the photon.
The simplest group with doublet representations is\footnote{
$SU(2)$ is the group of $2\times 2$ unitary matrices,
i.e. $U^\dagger U = U U^\dagger = 1$,  with $\det{U}=1$.
Any $SU(2)$ matrix can be written in the form
$U=\exp{\left\{i\vec{\alpha}\vec{\tau}/2\right\}}$, where
$\vec{\tau}$ are the usual Pauli matrices,
$$\tau_1=\left(\bat  0 & 1 \\ 1 & 0 \ea\right) , \qquad
  \tau_2=\left(\bat 0 & -i \\ i & 0 \ea\right) , \qquad
  \tau_3=\left(\bat 1 & 0 \\ 0 & -1 \ea\right),$$
which are traceless and satisfy the commutation relation
$
\left[ \tau_i,\tau_j\right]  =  2 i \epsilon_{ijk}\tau_k .
$
Other useful properties are:
$\left\{ \tau_i,\tau_j\right\}  =  2\delta_{ij}$
and $\mbox{\rm Tr}\left(\tau_i\tau_j\right)  =  2\delta_{ij}$.
}
 $SU(2)$.
We want to include also the electromagnetic interactions;
thus we need an additional $U(1)$ group. The obvious
symmetry group to consider is then
\bel{eq:group}
G\,\equiv\, SU(2)_L\otimes U(1)_Y ,
\ee
where $L$ refers to left-handed fields.
We do not specify, for the moment, the meaning of the
subindex $Y$ since, as we will see, the naive identification
with electromagnetism does not work.

For simplicity,
let us consider a single family of quarks, and introduce
the notation
\bel{eq:psi_def}
\psi_1(x)\, =\,\left(\ba u \\ d \ea\right)_L , \qquad
\psi_2(x) \, =\, u_R\, , \qquad\psi_3(x) \, = \, d_R \, .
\ee
Our discussion will also be valid for the
lepton sector, with the identification
\bel{eq:psi_def_l}
\psi_1(x)\, =\,\left(\ba \nu_e \\ e^- \ea\right)_L , \qquad
\psi_2(x) \, =\, (\nu_e)_R \, , \qquad\psi_3(x) \, = \, e^-_R\, .
\ee
%

As in the QED case, let us consider the free Lagrangian
\bel{eq:free_l}
\cL_0\, =\, \sum_{j=1}^3 \,i\,
\overline{\Psi}_j(x)\gamma^\mu\partial_\mu\Psi_j(x).
\ee
$\cL_0$ is invariant under
global
$G$ transformations,
\bel{eq:G_transf}
\psi_j(x)\,\toG\, \psi'_j(x)\,\equiv\,
\exp{\left\{i\vec{\tau}\vec{\alpha}/2\right\}}
\,\exp{\left\{iy_j\beta\right\}}\,\psi_j(x) ,
\ee
where the $SU(2)_L$ matrices only act on the doublet field
$\psi_1$.

We can now require the Lagrangian to be also invariant under
local gauge transformations $SU(2)_L\otimes U(1)_Y$,
i.e. with $\vec{\alpha}=\vec{\alpha}(x)$ and
$\beta=\beta(x)$.
In order to satisfy this symmetry requirement, we need to
change the fermion derivatives by covariant objects.
Since we have now 4 gauge parameters, $\vec{\alpha}(x)$
and $\beta(x)$, 4 different gauge bosons are needed:
\bel{eq:cov_der}
D_\mu\psi_j(x)\,\equiv\,\left[\partial_\mu
-ig\,{\vec{\tau}\over 2}\,\cdot\vecW_\mu
- ig'y_jB_\mu\right]\,\psi_j(x).
\ee
Thus, we have the correct number of gauge fields to describe
the $W^\pm$, $Z$ and $\gamma$.

We want $D_\mu\psi_j(x)$ to
transform in exactly the same way as the $\psi_j(x)$ fields;
this fixes the transformation properties of the gauge fields:
\beqn\label{eq:B_transf}
B_\mu(x)&\toG &B'_\mu(x)\,\equiv\, B_\mu(x) +{1\over g'}\,
\partial_\mu\beta(x) ,
\\ \label{eq:W-transf}
\vec{\tau}\,\cdot\vecW_\mu&\toG &
\vec{\tau}\,\cdot\vecWp_\mu\,\equiv\,
U(x)\,\vec{\tau}\,\cdot\vecW_\mu\, U^\dagger(x)
+ {2i\over g}\, U(x)\,\partial_\mu U^\dagger(x) ,
\eeqn
where
$U(x)\equiv\exp{\left\{i\vec{\tau}\vec{\alpha}(x)/2\right\}}$.
The transformation of $B_\mu$ is identical to the
one obtained in QED for the photon.
The $W^i_\mu$ fields transform in a more complicated way;
under an infinitesimal $SU(2)_L\otimes U(1)_Y$ transformation,
[$(a\times b)^i\equiv \epsilon_{ijk} a^j b^k$]
\bel{eq:W_transf_inf}
\vecWp_\mu\, = \, \vecW_\mu + {1\over g}\,\partial_\mu\vec{\alpha}
-\vec{\alpha}\,\times \vecW_\mu + \,\cO\left(\vec{\alpha}^2\right) .
\ee
The non-commutativity of the $SU(2)$ matrices gives rise to an
additional term $\vec{\alpha}\,\times\vecW_\mu$
involving the gauge fields themselves.
Note, that the $\psi_j$ couplings to the $B_\mu$ field
are completely free,
as in QED, i.e. there are arbitrary ``hypercharges'' $y_j$.
Since the $SU(2)$ commutation relation is non-linear,
this freedom does not exist for the $W^i_\mu$:
there  is only a unique $SU(2)_L$ coupling $g$.

The Lagrangian
\bel{eq:lagrangian}
\cL\, =\, \sum_{j=1}^3 \,i\,
\overline{\Psi}_j(x)\gamma^\mu D_\mu\Psi_j(x),
\ee
is invariant under local $G$ transformations.
In order to build the gauge-invariant kinetic term for the
gauge fields, we introduce the corresponding field strengths:
\bel{eq:b_mn}
B_{\mu\nu}  \,\equiv\,  \partial_\mu B_\nu - \partial_\nu B_\mu , \qquad\qquad
\vecW_{\mu\nu} \,\equiv\,
\partial_\mu \vecW_\nu - \partial_\nu \vecW_\mu
+ \, g \,\vecW_\mu\times \vecW_\nu .
\ee
$B_{\mu\nu}$ remains invariant under $G$ transformations,
while $\vec{\tau}\,\cdot\vecW_{\mu\nu}$ transforms covariantly:
\bel{eq:W_mn_transf}
\vec{\tau}\,\cdot\vecW_{\mu\nu}
\, = \, {2i\over g}\, \left[ \left(\partial_\mu - i g {\vec{\tau}\over 2}
\,\cdot \vecW_\mu\right)\, , \,\left(
\partial_\nu - i g {\vec{\tau}\,\over 2}\cdot \vecW_\nu\right)\right]
\,\toG\, U(x)\,\,
\vec{\tau}\,\cdot\vecW_{\mu\nu} \, U^\dagger(x) .
\ee
Therefore, the properly normalized kinetic Lagrangian is given by
\bel{eq:kinetic}
\cL_{\mbox{\rms Kin}} \, = \,
-{1\over 4} B_{\mu\nu} B^{\mu\nu} - {1\over 8}
\mbox{\rm Tr}\left[\left(\vec{\tau}\,\cdot\vecW_{\mu\nu}\right)
\left(\vec{\tau}\,\cdot\vecWmn\right)\right]
\, = \,
-{1\over 4} B_{\mu\nu} B^{\mu\nu} - {1\over 4}
\vecW_{\mu\nu}\vecWmn .
\ee
The non-abelian structure of the $SU(2)$ group generates here
an important difference with QED. Since the field strengths
$W^i_{\mu\nu}$ contain a quadratic piece, the Lagrangian
$\cL_{\mbox{\rms Kin}}$ gives rise to cubic and
quartic self-interactions among the gauge fields.
The strength of these interactions
is given by the same coupling $g$  which appears in the
fermionic piece of the Lagrangian.

The gauge symmetry forbids to write a mass term for the
gauge bosons. Fermionic masses are also not possible, because
they would communicate the left- and right-handed fields,
which have different transformation properties,
and therefore would produce an explicit
breaking of the gauge symmetry.
Thus, the $SU(2)_L\otimes U(1)_Y$ Lagrangian in
Eqs.~\eqn{eq:lagrangian} and \eqn{eq:kinetic}
only contains massless fields.

\subsection{Charged-current interaction}
\label{subsec:cc}

The Lagrangian \eqn{eq:lagrangian} contains interactions of the
fermion fields with the gauge bosons,
\bel{eq:int}
\cL\,\longrightarrow\,
{g\over 2}\,\overline{\Psi}_1\gamma^\mu (\vec{\tau}\,\cdot\vecW_\mu)
\Psi_1 + \, g'\,B_\mu\,
\sum_j\, y_j\,\overline{\Psi}_j\gamma^\mu\Psi_j \, .
\ee
The term containing the $SU(2)$ matrix
\bel{eq:W_matrix}
\vec{\tau}\,\cdot\vecW_\mu\, = \, \left(\bat
W^3_\mu & W_\mu^1-i W_\mu^2 \\ W_\mu^1+i W_\mu^2 & -W^3_\mu
\ea\right)
\ee
gives rise to charged-current interactions with the boson field
$W_\mu\equiv (W_\mu^1+i W_\mu^2)/\sqrt{2}$ and its complex-conjugate
$W^\dagger_\mu\equiv (W_\mu^1-i W_\mu^2)/\sqrt{2}$.
For a single family of quarks and leptons,
\bel{eq:W_interactions}
\cL_{\mbox{\rms CC}}\, = \, {g\over 2\sqrt{2}}\,\left\{
W^\dagger_\mu\,\left[
\bar u\gamma^\mu(1-\gamma_5) d + \bar\nu_e\gamma^\mu(1-\gamma_5) e
\right]\, + \, \mbox{\rm h.c.}\right\}\, .
\ee
Except for the missing $\theta_C$ mixing,
this is precisely the intermediate charged-boson interaction
assumed in Eq.~\eqn{eq:ivh}.
The universality of the quark and lepton interactions is now a direct
consequence of the gauge symmetry.
Note, however, that \eqn{eq:W_interactions} cannot
describe the observed dynamics, because the gauge boson is massless
and, therefore, gives rise to long-range forces.

\subsection{Neutral-current interaction}
\label{subsec:nc_int}

Eq.~\eqn{eq:int} contains also interactions with
the neutral gauge fields
$W^3_\mu$ and $B_\mu$. We would like to identify these bosons with
the $Z$ and the $\gamma$; but, since both fields are massless,
any arbitrary combination of them is a priori possible:
\bel{eq:Z_g_mixing}
\left(\ba W_\mu^3 \\ B_\mu\ea\right) \,\equiv\,
\left(\bat
\cos{\theta_W} & \sin{\theta_W} \\ -\sin{\theta_W} & \cos{\theta_W}
\ea\right) \, \left(\ba Z_\mu \\ A_\mu\ea\right)\, .
\ee
In terms of the fields $Z$ and $\gamma$,
the neutral-current Lagrangian is given by
\be\label{eq:L_NC}
\cL_{\mbox{\rms NC}} =
\sum_j\,\overline{\Psi}_j\gamma^\mu\left\{A_\mu
\left[{g\over 2}\tau_3\sin{\theta_W} + g'y_j\cos{\theta_W}\right]
 +
Z_\mu
\left[{g\over 2}\tau_3\cos{\theta_W} - g'y_j\sin{\theta_W}\right]
\right\}\Psi_j\, .
\ee
%
%
In order to get QED from the $A_\mu$ piece, one needs to impose the
conditions:
\bel{eq:unification}
g\,\sin{\theta_W} \, = \, g'\,\cos{\theta_W}\, = \, e  ,
\qquad\qquad\qquad
 Y \, = \, Q-T_3  ,
\ee
where $T_3\equiv\tau_3/2$ and
$Q$ denotes the electromagnetic charge operator
\bel{eq:Q}
Q_1\,\equiv\,\left(\bat Q_{u/\nu}& 0 \\ 0 & Q_{d/e}\ea\right)\, ,
\qquad Q_2\, =\,Q_{u/\nu}\, ,\qquad Q_3\, =\,Q_{d/e}
\, .
\ee
The first equality relates the $SU(2)_L$ and $U(1)_Y$ couplings
to the electromagnetic coupling, providing the wanted unification
of the electroweak interactions. The second identity, fixes
the fermion hypercharges in terms of their electric charge and
weak isospin quantum numbers:
$y_1=Q_{u/\nu}-1/2=Q_{d/e}+1/2$, $y_2=Q_{u/\nu}$ and
$y_3=Q_{d/e}$.
Note that a hypothetical
right-handed neutrino would have both electric charge
and weak hypercharge equal to zero; since it would not couple
either to the $W$ boson, such a particle would not have any
kind of interaction (sterile neutrino).
For aesthetical reasons, we will then not consider right-handed
neutrinos any longer.

\begin{table}[tbh]
\begin{center}
\caption{Neutral-current couplings.\label{tab:nc_couplings}}
\vspace{0.2cm}
\begin{tabular}{|c||c|c||c|c|}
\hline
& $u$ & $d$ & $\nu_e$ & $e$
\\ \hline\hline
$\, v_f\, $ & $(1 -{8\over 3} \sin^2{\theta_W})/2$ &
$(-1 +{4\over 3} \sin^2{\theta_W})/2$ & $\,\, 1/2\,\,$ &
$(-1 +4 \sin^2{\theta_W})/2$
\\
$a_f$ & $1/2$ & $-1/2$ & $1/2$ & $-1/2$
\\ \hline
\end{tabular}
\end{center}
\end{table}

Using the relations \eqn{eq:unification},
the neutral-current Lagrangian can be written as
\bel{eq:L_NCb}
\cL_{\mbox{\rms NC}}\, = \,
\cL_{\mbox{\rms QED}}\, + \,
\cL_{\mbox{\rms NC}}^Z\, ,
\ee
where
\bel{eq:L_QED}
\cL_{\mbox{\rms QED}}\, = \,e\,A_\mu\,\sum_j\,
\overline{\Psi}_j\gamma^\mu Q_j\Psi_j
\,\equiv\, e\, A_\mu\, J^\mu_{\mbox{\rms em}}
\ee
is the usual QED Lagrangian and
\beqn\label{eq:Z_NC}
& &\qquad
\cL_{\mbox{\rms NC}}^Z \, =\,
{e\over 2 \sin{\theta_W}\cos{\theta_W}} \,
J^\mu_Z\, Z_\mu\, ,
\\
J^\mu_Z &\!\! \equiv &\!\! \sum_j\,\overline{\Psi}_j\gamma^\mu
\left(\tau_3-2\sin^2{\theta_W}Q_j\right)\Psi_j
\, = \, J^\mu_3 - 2 \sin^2{\theta_W}\, J^\mu_{\mbox{\rms em}}\, ,
\eeqn
contains the $Z$-boson interactions.
In terms of the more usual fermion fields,
$\cL_{\mbox{\rms NC}}^Z$ has the form
\bel{eq:Z_Lagrangian}
\cL_{\mbox{\rms NC}}^Z\, = \,
{e\over 2\sin{\theta_W}\cos{\theta_W}} \,Z_\mu\,\sum_f\,
\bar f \gamma^\mu (v_f-a_f\gamma_5) \, f\, ,
\ee
where
$a_f =  T_3^f$ and
$v_f = T_3^f \left( 1 - 4 |Q_f| \sin^2{\theta_W}\right)$.

\subsection{Gauge self-interactions}
\label{subsec:self-interactions}

In addition to the usual kinetic terms, the Lagrangian
\eqn{eq:kinetic} generates cubic and quartic self-interactions
among the gauge bosons:
\beqn\label{eq:cubic}
\cL_3 &\!\!\!\! = &\!\!\!\!
-i e \cot{\theta_W}\left\{
\left(\partial^\mu W^\nu -\partial^\nu W^\mu\right)
 W^\dagger_\mu Z_\nu -
\left(\partial^\mu W^{\nu\dagger} -\partial^\nu W^{\mu\dagger}\right)
 W_\mu Z_\nu +
W_\mu W^\dagger_\nu\left(\partial^\mu Z^\nu -\partial^\nu Z^\mu\right)
\right\}
\no\\ &&\!\!\!\!\!\!\!\!\!\!
-i e \left\{
\left(\partial^\mu W^\nu -\partial^\nu W^\mu\right)
 W^\dagger_\mu A_\nu -
\left(\partial^\mu W^{\nu\dagger} -\partial^\nu W^{\mu\dagger}\right)
 W_\mu A_\nu +
W_\mu W^\dagger_\nu\left(\partial^\mu A^\nu -\partial^\nu A^\mu\right)
\right\}  ;
\\ \label{eq:quartic}
\cL_4 &\!\!\!\! = &\!\!\!\!
-{e^2\over 2\sin^2{\theta_W}}\left\{
\left(W^\dagger_\mu W^\mu\right)^2 - W^\dagger_\mu W^{\mu\dagger}
W_\nu W^\nu \right\}
- e^2 \cot^2{\theta_W}\,\left\{
W_\mu^\dagger W^\mu Z_\nu Z^\nu - W^\dagger_\mu Z^\mu W_\nu Z^\nu
\right\}
\no\\ &\!\!\!\! &\!\!\!\!\!\!\!\!\!\!
- e^2 \cot{\theta_W}\left\{
2 W_\mu^\dagger W^\mu Z_\nu A^\nu - W^\dagger_\mu Z^\mu W_\nu A^\nu
- W^\dagger_\mu A^\mu W_\nu Z^\nu
\right\}
\no\\ &\!\!\!\! &\!\!\!\!\!\!\!\!\!\!
- e^2\,\left\{
W_\mu^\dagger W^\mu A_\nu A^\nu - W^\dagger_\mu A^\mu W_\nu A^\nu
\right\} .
\eeqn
Notice 
that $\cL_3$ has only terms with two
charged $W$'s and one neutral ($Z$ or $\gamma$) boson.

\setcounter{equation}{0}
\section{Spontaneous Symmetry Breaking}
\label{sec:ssb}

So far, we have been able to derive charged- and neutral-current
interactions of the type needed to describe weak decays;
we have nicely incorporated QED into the
same theoretical framework; and, moreover, we have got additional
self-interactions of the gauge bosons, which are generated by the
non-abelian structure of the $SU(2)$ group.
Gauge symmetry also guarantees that we have a well-defined
renormalizable Lagrangian.
However, this Lagrangian has very little to do with
reality. Our gauge bosons are massless particles; while this is
fine for the photon field, the physical $W^\pm$ and
$Z$ bosons should be quite heavy objects.

In order to generate masses, we need to break the gauge symmetry
in some way; however, we also need a fully symmetric Lagrangian
to preserve renormalizability.
A possible solution to this dilemma, is based on the fact that
it is possible to get non-symmetric results from an invariant
Lagrangian.

Let us consider a Lagrangian, which:
\begin{enumerate}
\item Is invariant under a group $G$ of transformations.
\item Has a degenerate set of states with minimal energy,
which transform under $G$ as the members of a given multiplet.
\end{enumerate}
If one arbitrarily selects one of these states as the ground
state of the system, one says that the symmetry becomes
spontaneously broken.

This kind of situation is clearly illustrated by the so-called
Buridan's donkey dilemma:  
imagine a donkey at equal distance from two equal amounts of food;
while this is a perfectly symmetric scenario, the symmetry will
be ``spontaneously'' broken when the donkey will decide which one
it is going to eat first.
A more physical example is provided by a ferromagnet: although
the Hamiltonian is invariant under rotations, the ground state
has the spins aligned into some arbitrary direction. Moreover,
any higher-energy state, built from the ground state by a finite
number of excitations, would share its anisotropy.

In a Quantum Field Theory, the ground state is the vacuum.
Thus, the SSB mechanism will appear in those cases where
one has a symmetric Lagrangian, but a non-symmetric vacuum.

\subsection{Goldstone theorem}
\label{subsec:goldstone}

Let us consider a complex scalar field $\phi(x)$, with
Lagrangian
\bel{eq:L_phi}
\cL\, = \, \partial_\mu\phi^\dagger \partial^\mu\phi - V(\phi) ;
\qquad\qquad
V(\phi)\, = \, \mu^2 \phi^\dagger\phi + h
\left(\phi^\dagger\phi\right)^2 .
\ee
$\cL$ is invariant under global phase transformations
of the scalar field
\bel{eq:phi_transf}
\phi(x)\,\longrightarrow\,\phi'(x)\,\equiv\,
\exp{\left\{i\theta\right\}}\,\phi(x) \, .
\ee

In order to have a ground state the potential should be bounded
from below, i.e. $h>0$. For the quadratic piece there are two
possibilities:
\begin{enumerate}
\item \mbox{\boldmath $\mu^2>0$:}
The potential has only the trivial minimum $\phi=0$.
It describes a massive scalar particle with mass $\mu$
and quartic coupling $h$.
\item \mbox{\boldmath $\mu^2<0$:}
The minimum is obtained for those field configurations
satisfying
\bel{eq:minimum}
|\phi_0|\, = \, \sqrt{{-\mu^2\over 2 h}} \,\equiv\, {v\over\sqrt{2}}
\, > \, 0 ;
\qquad\qquad\qquad V(\phi_0)\, =\, -{h\over 4} v^4 .
\ee
Owing to the $U(1)$ phase-invariance of the Lagrangian,
there is an infinite number of degenerate states of
minimum energy,
$\phi_0(x) = {v\over\sqrt{2}} \exp{\left\{i\theta\right\}}$.
By choosing a particular solution, $\theta=0$ for example, as the
ground state, the symmetry gets spontaneously broken.
If we parametrize the excitations over the ground state as
\bel{eq:perturbations}
\phi(x)\, \equiv \, {1\over\sqrt{2}}\,\left[
v + \phi_1(x) + i \phi_2(x)\right] ,
\ee
where $\phi_1$ and $\phi_2$ are real fields,
the potential takes the form
\bel{eq:pot}
V(\phi)\, = \, V(\phi_0) -\mu^2\phi_1^2 +
h v \phi_1 \left(\phi_1^2+\phi_2^2\right) +
{h\over 4} \left(\phi_1^2+\phi_2^2\right)^2 .
\ee
Thus, $\phi_1$ describes a massive state of mass
$m_{\phi_1}^2 = -2\mu^2$, while $\phi_2$ is massless.
\end{enumerate}

The first possibility ($\mu^2>0$) is just the usual situation
with a single ground state.
The other case, with SSB, is more interesting.
The appearence of a massless particle when $\mu^2<0$
is easy to understand: the field $\phi_2$ describes excitations
around a flat direction in the potential, i.e. into states
with the same energy as the chosen ground state.
Since those excitations do not cost any energy, they obviously
correspond to a massless state.

The fact that there are massless excitations associated with the
SSB mechanism is a completely general result, known
as the Goldstone theorem \cite{goldstone}:
if a Lagrangian is invariant under a continuous symmetry group
$G$, but the vacuum is only invariant under a subgroup $H\subset G$,
then there must exist as many massless spin--0 particles
(Goldstone bosons) as broken generators
(i.e. generators of $G$ which do not belong to $H$).

\subsection{The Higgs--Kibble mechanism}
\label{subsec:Higgs-Kibble}

At first sight, the Goldstone theorem has very little to do with
our mass problem; in fact, it makes it worse since we want
massive states and not massless ones.
However, something very interesting happens when there is a local
gauge symmetry \cite{HI:66}.

Let us consider \cite{WE:67}
an $SU(2)_L$ doublet of complex scalar fields
\bel{eq:scalar_multiplet}
\phi(x)\,\equiv\,\left(\ba \phi^{(+)}(x)\\ \phi^{(0)}(x)\ea\right) .
\ee
The gauged scalar Lagrangian of the Goldstone model (\ref{eq:L_phi}),
\beqn\label{eq:LS}
&&\!\!\cL_S \, = \, \left(D_\mu\phi\right)^\dagger D^\mu\phi
-\mu^2\phi^\dagger\phi - h \left( \phi^\dagger\phi\right)^2 ,
\qquad\qquad (h>0,\, \mu^2<0) ,
\\ \label{eq:DS}
&&\!\!D^\mu\phi \, = \,
\left[\partial_\mu
-ig\,{\vec{\tau}\over 2}\,\cdot\vecW_\mu
- ig'y_\phi B_\mu\right]\,\phi , \qquad\qquad\quad (y_\phi=1/2) ,
\eeqn
is invariant under local $SU(2)_L\otimes U(1)_Y$ transformations.
The value of the scalar hypercharge is fixed by the requirement
of having the correct couplings between $\phi(x)$ and
$A^\mu(x)$; i.e. that the photon does not couple to $\phi^{(0)}$,
and one has the right electric charge for $\phi^{(+)}$.

The potential is very similar to the one considered before.
There is a infinite set of degenerate states with minimum
energy, satisfying
\bel{eq:vev}
\big|\langle 0|\phi^{(0)}|0\rangle\big|
\, = \, \sqrt{{-\mu^2\over 2 h}} \,\equiv\, {v\over\sqrt{2}} \, .
\ee
Note that we have made explicit the association of the classical
ground state with the quantum vacuum. Since the electric charge
is a conserved quantity, only the neutral scalar field can acquire
a vacuum expectation value.
Once we choose a particular ground state,
the $SU(2)_L\otimes U(1)_Y$ symmetry gets spontaneously broken to
the electromagnetic subgroup $U(1)_{\mbox{\rms QED}}$,
which by construction still remains a true symmetry of the vacuum.
According to Goldstone theorem 3 massless states should then appear.

Now, let us parametrize the scalar doublet in the general form
\bel{eq:parametrization}
\phi(x)\, = \, \exp{\left\{
i {\vec{\tau}\over 2}\,\cdot\vec{\theta}(x)\right\}}
\, {1\over\sqrt{2}}\,
\left(\ba 0 \\ v + H(x) \ea\right) ,
\ee
with 4 real fields $\vec{\theta}(x)$ and $H(x)$.
The crucial point to be realized is that the local $SU(2)_L$
invariance of the Lagrangian allows us to rotate away any
dependence on $\vec{\theta}(x)$.
These 3 fields are precisely the would-be massless Goldstone bosons
associated with the SSB  mechanism.
The additional ingredient of gauge symmetry makes those
massless excitations unphysical.

The covariant derivative \eqn{eq:DS} couples the scalar
multiplet to the $SU(2)_L\otimes U(1)_Y$ gauge bosons.
If one takes the physical (unitary) gauge $\vec{\theta}(x)=\vec{0}$,
the kinetic piece of the scalar Lagrangian \eqn{eq:LS}
takes the form:
\bel{eq:unitary_gauge}
\left(D_\mu\phi\right)^\dagger D^\mu\phi\,\, \toTheta\,\,
{1\over 2} \partial_\mu H \partial^\mu H
+ (v+H)^2 \, \left\{ {g^2\over 4} W_\mu^\dagger W^\mu
+ {g^2\over 8\cos^2{\theta_W}} Z_\mu Z^\mu \right\} .
\ee
The vacuum expectation value of the neutral scalar has
generated a quadratic term for the $W^\pm$ and the $Z$,
i.e. those gauge bosons have acquired masses:
\bel{eq:boson_masses}
M_Z\,\cos{\theta_W}\, = \, M_W \, = \,  v g/2 \, .
\ee

Therefore, we have found a clever way of giving masses to the
intermediate carriers of the weak force. We just add $\cL_S$
to our $SU(2)_L\otimes U(1)_Y$ model.
The total Lagrangian is invariant under gauge transformations,
which guarantees \cite{TH:71}
the renormalizability of the associated
Quantum Field Theory. However, SSB occurs. The 3 broken generators
give rise to 3 massless Goldstone bosons which,
owing to the underlying local gauge symmetry,
are unphysical (i.e. do not produce any observable effect).
Going to the unitary gauge, we discover that the $W^\pm$ and the
$Z$ (but not the $\gamma$, because $U(1)_{\mbox{\rms QED}}$ is
an unbroken symmetry) have acquired masses, which are moreover related
as indicated in Eq.~\eqn{eq:boson_masses}.
Notice that (\ref{eq:Z_g_mixing}) has now the meaning of writing
the gauge fields in terms of the physical boson fields with
definite mass.

It is instructive to count the number of degrees of freedom (d.o.f.).
Before the SSB mechanism, the Lagrangian contains massless
$W^\pm$ and $Z$ bosons (i.e. $3\times 2 = 6$ d.o.f., due to the 2
possible polarizations of a massless spin--1 field) and 4
real scalar fields. After SSB, the 3 Goldstone modes are ``eaten''
by the weak gauge bosons, which become massive and, therefore,
acquire one additional longitudinal polarization.
We have then $3\times 3=9$ d.o.f. in the gauge sector,
plus the remaining
scalar particle $H$, which is called the Higgs boson. The total
number of d.o.f. remains of course the same.

\subsection{Predictions}
\label{subsec:predictions}

We have now all the needed ingredients to describe weak interactions.
We can reproduce the old low-energy results mentioned
in Sect.~\ref{sec:history},
within a well-defined Quantum Field Theory.
Our theoretical framework predicts the existence of massive
intermediate gauge bosons, $W^\pm$ and $Z$, which
have been confirmed \cite{cern:83} by
the modern high-energy colliders.
Moreover,
the Higgs-Kibble mechanism has produced a precise
prediction\footnote{
Note, however, that the relation
$M_Z\cos{\theta_W} =  M_W$ has a more general validity.
It is a direct consequence of the symmetry properties of
$\cL_S$ and does not depend on its detailed dynamics.}
for
the $W^\pm$ and $Z$ masses, relating them to the vacuum expectation
value of the scalar field through Eq.~\eqn{eq:boson_masses}.
Thus, $M_Z$ is predicted to be bigger than $M_W$.
Using the relations
$G_F/\sqrt{2}=g^2/(8M_W^2)$ and $e=g\sin{\theta_W}$, we get
\bel{eq:M_W}
M_W  =  \left( {\pi\alpha\over G_F\sqrt{2}}\right)^{1/2}\,
{1\over\sin{\theta_W}} =
{37.280\,\mbox{\rm GeV}\over\sin{\theta_W}}\,  ,
\qquad\quad
v  =  \left(\sqrt{2} G_F\right)^{-1/2} = 246\,\mbox{\rm GeV}\, .
\ee

A direct test of these relations can be obtained in
neutrino-scattering experiments, by comparing the cross-sections
of neutral-current and charged-current processes.
The elastic scattering $\nu q\to\nu q$ occurs through $Z$-exchange
in the $t$ channel, whereas the inelastic process
$\nu q\to l q'$ requires the exchange of a charged $W$.
At low momentum transfer the boson propagators reduce to
constants, given by the
corresponding masses;
moreover, the fermionic couplings of the $Z$ and
the $W^\pm$ in Eqs.~\eqn{eq:Z_Lagrangian} and \eqn{eq:W_interactions}
are related by the weak mixing angle $\theta_W$.
Therefore,
\bel{eq:NC_CC}
{\sigma_{\mbox{\rms NC}}(\nu q)\over\sigma_{\mbox{\rms CC}}(\nu q)}
\,\sim\,
\left({M_W^2\over M_Z^2\cos^2{\theta_W}}\right)^2\,
f\left(\sin^2{\theta_W}\right) .
\ee
One can, moreover, compare $\nu$ and $\bar\nu$ scattering processes
on different targets.
The analysis of the experimental data gives \cite{pdg:94}
\bel{eq:exp}
{M_W^2\over M_Z^2\cos^2{\theta_W}}\,\approx\, 1 ;
\qquad\qquad\qquad
\sin^2{\theta_W}\,\approx\, 0.23 \, .
\ee
The excellent agreement with the theoretical prediction
constitutes a very succesful confirmation of the assumed pattern
of SSB.
Inserting the measured value of $\theta_W$ in Eq.~\eqn{eq:M_W},
one gets numerical predictions for the gauge-boson masses,
\bel{eq:th_masses}
M_W\, \approx\, 78\,\mbox{\rm GeV} ,
\qquad\qquad\qquad
M_Z\, \approx\, 89\,\mbox{\rm GeV} ,
\ee
which are in quite good agreement with the experimental measurements,
$M_W=  (80.23\pm 0.18)\,\mbox{\rm GeV}$  and
$M_Z = (91.1888\pm 0.0044)\,\mbox{\rm GeV}$ \cite{lep:94,MW:94}.
The small numerical discrepancies can be understood in terms
of higher-order quantum corrections (see Sects. \ref{sec:tree}
and \ref{sec:nc-loop}).

\subsection{The Higgs boson}
\label{subsec:Higgs}

The scalar Lagrangian \eqn{eq:LS}
has introduced a new scalar particle into the model: the Higgs $H$.
In terms of the physical fields (unitary gauge), $\cL_S$ takes the
form
\bel{eq:H_lag}
\cL_S\, = \, {h v^4\over 4} \, + \,
\cL_H \, + \, \cL_{HG^2} \, ,
\ee
where
\beqn\label{eq:H_int}
&&\cL_H \, = \, {1\over 2} \partial_\mu H \partial^\mu H -
{1\over 2} M_H^2 H^2
- {M_H^2\over 2 v} H^3 - {M_H^2\over 8 v^2} H^4 ,
\\ \label{eq:HGG}
\cL_{HG^2} &\! = &\! M_W^2\, W_\mu^\dagger W^\mu\,
\left\{ 1+ {2\over v} H
+ {H^2\over v^2}\right\}\, + \,
{1\over 2} M_Z^2\, Z_\mu Z^\mu \,\left\{ 1+{2\over v} H
+ {H^2\over v^2}\right\} ,
\eeqn
and the Higgs mass is given by
\bel{H_mass}
M_H\, = \, \sqrt{-2\mu^2}\, =\, \sqrt{2 h}\, v \, .
\ee
Notice that the Higgs interactions have a very characteristic form:
they are always proportional to the mass (squared) of the coupled boson.
All Higgs couplings
are determined by $M_H$, $M_W$, $M_Z$ and the vacuum expectation value $v$.

\setcounter{equation}{0}
\section{Anomalies}
\label{sec:anomalies}

Our theoretical framework is based on the local gauge symmetry.
However, we have only
discussed so far the symmetries of the classical Lagrangian.
It happens sometimes that a symmetry of $\cL$ gets broken by quantum effects,
i.e. it is not a symmetry of the quantized theory;
one says then that there is an ``anomaly''.

Anomalies appear in those symmetries involving both axial
($\overline{\Psi}\gamma^\mu\gamma_5\Psi$) and vector
($\overline{\Psi}\gamma^\mu\Psi$)
currents, and reflect the impossibility of regularizing the quantum
theory (the divergent loops) in a way which preserves the chiral
(left/right) symmetries.

A priori there is nothing wrong with having an anomaly. In fact, sometimes
they are even welcome. A good example is provided by the decay
$\pi^0\to\gamma\gamma$. There is a (chiral) symmetry of the QCD Lagrangian
which forbids this transition; the $\pi^0$ should then be a stable particle,
in contradiction with the experimental evidence.
Fortunately, there is an anomaly generated by a triangular quark loop
which couples the axial current
$A_\mu^3\equiv (\bar u\gamma_\mu\gamma_5 u - \bar d\gamma_\mu\gamma_5 d)$
to two electromagnetic currents
and breaks the conservation of the axial current at the quantum level:
\bel{eq:div_A}
\partial^\mu A_\mu^3 \,=\,  {\alpha\over 4\pi}\,
\epsilon^{\alpha\beta\sigma\rho}\,
F_{\alpha\beta}
\, F_{\sigma\rho} \, + \, \cO\left(m_u + m_d\right) .
\ee
Since the $\pi^0$ couples to $A_\mu^3$,
the $\pi^0\to\gamma\gamma$
decay does finally occur, with a predicted
rate
\bel{eq:pi_decay}
\Gamma(\pi^0\to\gamma\gamma)\, = \, \left({N_C\over 3}\right)^2
{\alpha^2 m_\pi^3\over 64\pi^3 f_\pi^2}
\, =\, 7.73\, \mbox{\rm eV} ,
\ee
where $N_C=3$ denotes the number of quark ``colours''.
The agreement with the measured value, $\Gamma = 7.7\pm 0.6$ eV
\cite{pdg:94}, is excellent.

Anomalies are, however, very dangerous in the case of local gauge symmetries,
because they destroy the renormalizability of the Quantum Field Theory.
Since the $SU(2)_L\otimes U(1)_Y$ model is chiral (i.e. it distinguishes
left from right), anomalies are clearly present.
The gauge bosons couple to vector and axial-vector currents; we can then
draw triangular diagrams
with
three arbitrary gauge bosons ($W^\pm$, $Z$, $\gamma$) in the external legs.
Any such diagram involving one axial and two vector currents generates
a breaking of the gauge symmetry.
Thus, our nice model looks meaningless at the quantum level.

We have still one way out. What matters is not the value of a single
Feynman diagram, but the sum of all possible contributions.
The anomaly generated by the sum of all triangular diagrams
connecting the three gauge bosons $G_a$, $G_b$ and $G_c$ is proportional
to
\bel{eq:an_condition}
\cA\, = \, \mbox{\rm Tr}\left( \{ T^a , T^b \} T^c \right)_L -
 \mbox{\rm Tr}\left( \{ T^a , T^b \} T^c \right)_R,
\ee
where the traces sum over all possible left- and right-handed
fermions, respectively, running along the internal lines
of the triangle.
The matrices $T^a$ are the generators associated with the corresponding
gauge bosons; in our case, $T^a = \tau_a/2, Y$.

In order to preserve the gauge symmetry, one needs a cancellation of all
anomalous contributions, i.e. $\cA=0$.
Since $\mbox{\rm Tr}(\tau_k)=0$, we have an automatic cancellation in two
combinations of generators:
$\mbox{\rm Tr}\left(\{ \tau_i , \tau_j \} \tau_k \right)=2\delta^{ij}
\mbox{\rm Tr}(\tau_k)=0$,
$\mbox{\rm Tr}\left(\{ Y , Y\} \tau_k \right)\propto\mbox{\rm Tr}(\tau_k)= 0$.
However, the other two combinations,
$\mbox{\rm Tr}\left(\{ \tau_i , \tau_j \} Y \right)$ and
$\mbox{\rm Tr}(Y^3)$ turn out to be proportional to
$\mbox{\rm Tr}(Q)$, i.e. to the sum of fermion electric charges:
\bel{eq:an_cancellation}
\sum_i Q_i \, = \, Q_e + Q_\nu + N_C \left( Q_u + Q_d \right) \, = \,
-1 + {1\over 3} N_C .
\ee

Eq.~\eqn{eq:an_cancellation} is telling us a very important message: the gauge
symmetry of the $SU(2)_L\otimes U(1)_Y$ model does not have any quantum
anomaly, provided that $N_C=3$.
Fortunately, this is precisely the right number of colours to understand
strong interactions.
Thus, at the quantum level, the electroweak model seems to know something
about QCD. The complete SM gauge theory based on the group
$SU(3)_C\otimes SU(2)_L\otimes U(1)_Y$ is free of anomalies and, therefore,
renormalizable.

\setcounter{equation}{0}
\section{Fermion Generations}
\label{sec:flavour}

\subsection{The GIM mechanism}
\label{subsec:GIM}

The $V-A$ low-energy Hamiltonian \eqn{eq:v_a} shows that the
$SU(2)_L$ partner of the up quark should not be the $d$, but rather
the combination $d_C=\cos{\theta_C} d + \sin{\theta_C} s$.
However, if one naively replaces $d$ by $d_C$ in the neutral-current
Lagrangian \eqn{eq:Z_Lagrangian}, one generates a flavour-changing
neutral-current coupling,
\bel{eq:FCNC_disaster}
Z_\mu\, \bar d_C \gamma^\mu (v_d-a_d\gamma_5) d_C \,\longrightarrow\,
\cos{\theta_C} \sin{\theta_C}\,
Z_\mu\, \left[\bar d \gamma^\mu (v_d-a_d\gamma_5) s
\, + \, \bar s \gamma^\mu (v_d-a_d\gamma_5) d \right] ,
\ee
of a similar magnitude than the flavour-conserving $Z\bar d d$ one.
This is a major phenomenological disaster, in view of the strong
experimental bounds in Eq.~\eqn{eq:fcnc}.

In order to solve this problem, it was suggested in 1970 \cite{GIM:70}
that an additional quark flavour should exist: the charm.
One could then form two different quark doublets,
\bel{eq:two_doublets}
\left(\ba u \\ d_C\ea\right) , \qquad\qquad
\left(\ba c \\ s_C\ea\right) ,
\ee
with
\bel{eq:c_mixing}
\left(\ba d_C \\ s_C\ea\right)\, = \,
\left(\bat \cos{\theta_C} &\sin{\theta_C} \\ -\sin{\theta_C}& \cos{\theta_C}\ea
\right)\, \left(\ba d \\ s\ea\right)\,
\equiv\, \mbox{\boldmath $V$}\, \left(\ba d \\ s\ea\right) .
\ee
The orthogonality of the quark-mixing matrix {\boldmath $V$}
would then preserve the required absence of flavour-changing neutral
couplings (GIM mechanism \cite{GIM:70}),
\bel{eq:gim_mechanism}
\left(\bar d_C d_C + \bar s_C s_C\right)
\, =\, \left(\bar d d + \bar s s\right) ,
\ee
as long as the couplings of the two doublets are identical.
The discovery of the charm quark in 1974 \cite{charm} was a big step
forward in the development of the SM.

\subsection{Fermion masses}
\label{subsec:f_masses}

In order to properly speak about quark flavours, we need first to
understand the quark masses
($d$ and $s$ are defined as mass-eigenstates).
%
%
We know already that a fermionic mass term,
$\cL_m = -m \overline{\Psi}\Psi = - m \left(\overline{\Psi}_L\Psi_R
+  \overline{\Psi}_R\Psi_L\right)$
is not allowed, because it breaks the gauge symmetry.
However, since we have introduced an additional scalar doublet into
the model, we can write the following gauge-invariant fermion-scalar coupling:
\bel{eq:yukawa}
\cL_Y\, =\, c_1\, \left(\bar u , \bar d\right)_L
\left(\ba \phi^{(+)}\\ \phi^{(0)}\ea\right)\, d_R \, + \,
c_2\,\left(\bar u , \bar d\right)_L
\left(\ba \phi^{(0)\dagger}\\ -\phi^{(+)\dagger}\ea\right)\, u_R \, + \,
c_3\,\left(\bar \nu_e , \bar e\right)_L
\left(\ba \phi^{(+)}\\ \phi^{(0)}\ea\right)\, e_R
\, +\, \mbox{\rm h.c.}
\ee
In the unitary gauge (after SSB),
this Yukawa-type Lagrangian takes the simpler form
\bel{eq:y_m}
\cL_Y\, =\, {1\over\sqrt{2}} \,(v+H)\,\left\{ c_1 \,\bar d d + c_2
\,\bar u u + c_3 \,\bar e e\right\} .
\ee
Therefore, the SSB mechanism 
also generates fermion masses:
\bel{eq:f_masses}
m_d\, = \, - c_1 {v/\sqrt{2}} \, ; \qquad
m_u\, = \, - c_2 {v/\sqrt{2}} \, ; \qquad
m_e\, = \, - c_3 {v/\sqrt{2}} \, .
\ee

Since we do not know the parameters $c_i$, the values of the fermion masses
are arbitrary. Note, however, that all Yukawa couplings are fixed in terms
of the masses:
\bel{eq:y_mass}
\cL_Y\, =\, -\left( 1 + {H\over v}\right)
 \,\left\{ m_d \,\bar d d + m_u \,\bar u u
+ m_e \,\bar e e\right\} .
\ee

\subsection{Flavour mixing}
\label{subsec:FlavourMixing}

We have learnt experimentally that there are 6 different quark flavours
($u$, $d$, $s$, $c$, $b$, $t$), 3 different leptons ($e$, $\mu$, $\tau$)
and their corresponding neutrinos ($\nu_e$, $\nu_\mu$, $\nu_\tau$).
We can nicely include all these particles into the SM framework,
by organizing them into 3 families of quarks and leptons, as
indicated in Eqs. \eqn{eq:families} and \eqn{eq:structure}.
Thus, we have 3 nearly-identical copies of the same
$SU(2)_L\otimes U(1)_Y$ structure, with masses as the only difference.

Let us consider the general case of $N_G$ generations of fermions,
and denote $\nu_j'$, $l'_j$, $u'_j$, $d'_j$ the members of the weak
family $j$ ($j=1,\ldots,N_G$),
with definite transformation properties under the gauge group.
The weak eigenstates are linear
combinations of mass eigenstates.
The most general Yukawa Lagrangian has the form
\beqn\label{eq:N_Yukawa}
\cL_Y &=&\sum_{jk}\,\left\{
\left(\bar u'_j , \bar d'_j\right)_L \left[ c^{(d)}_{jk}
\left(\ba \phi^{(+)}\\ \phi^{(0)}\ea\right)\, d'_{kR}  +
c^{(u)}_{jk}
\left(\ba \phi^{(0)\dagger}\\ -\phi^{(+)\dagger}\ea\right)\, u'_{kR}
\right]
\right.\no\\ && \quad\,\,\,\left. +
\left(\bar \nu'_j , \bar l'_j\right)_L c^{(l)}_{jk}
\left(\ba \phi^{(+)}\\ \phi^{(0)}\ea\right)\, l'_{kR}\right\}
\, +\, \mbox{\rm h.c.},
\eeqn
where $c^{(d)}_{jk}$, $c^{(u)}_{jk}$ and $c^{(l)}_{jk}$
are arbitrary coupling constants.

After SSB, the Yukawa Lagrangian can be written as
\bel{eq:N_Yuka}
\cL_Y = - \left(1 + {H\over v}\right)\,\left\{
\overline{\mbox{\boldmath $d$}}'_L \mbox{\boldmath $M$}_d'
\mbox{\boldmath $d$}'_R \, + \,
\overline{\mbox{\boldmath $u$}}'_L \mbox{\boldmath $M$}_u'
\mbox{\boldmath $u$}'_R
\, + \,
\overline{\mbox{\boldmath $l$}}'_L \mbox{\boldmath $M$}'_l
\mbox{\boldmath $l$}'_R \, +
\mbox{\rm h.c.}\right\} .
\ee
Here, $\mbox{\boldmath $d$}'$, $\mbox{\boldmath $u$}'$
and $\mbox{\boldmath $l$}'$ denote vectors in flavour
space, and the corresponding mass matrices are given by
\bel{eq:M_c_relation}
(\mbox{\boldmath $M$}'_d)_{ij}\,\equiv\,
- c^{(d)}_{ij}\, v/\sqrt{2}\, , \qquad
(\mbox{\boldmath $M$}'_u)_{ij}\,\equiv\,
- c^{(u)}_{ij}\, v/\sqrt{2}\, , \qquad
(\mbox{\boldmath $M$}'_l)_{ij}\,\equiv\,
 - c^{(l)}_{ij}\, v/\sqrt{2}\, .
\ee
The diagonalizacion of these mass matrices determines the mass
eigenstates $d_j$, $u_j$ and $l_j$.

The matrix $\mbox{\boldmath $M$}_d'$ can be decomposed as\footnote{
The condition $\det{\mbox{\protect\boldmath $M$}'_f}\not=0$
($f=d,u,l$)
guarantees that the decomposition
$\mbox{\protect\boldmath $M$}'_f=
\mbox{\protect\boldmath $H$}_f
\mbox{\protect\boldmath $U$}_f$ is unique:
$\mbox{\protect\boldmath $U$}_f\equiv
\mbox{\protect\boldmath $H$}_f^{-1}\mbox{\protect\boldmath $M$}_f'$.
The matrices $\mbox{\protect\boldmath $S$}_f$
are completely determined (up to phases)
only if all diagonal elements of $\mbox{\protect\boldmath $\cM$}_f$
are different.
If there is some degeneracy, the arbitrariness of
$\mbox{\protect\boldmath $S$}_f$
reflects the freedom to define the physical fields.
If $\det{\mbox{\protect\boldmath $M$}'_f}=0$,
the matrices $\mbox{\protect\boldmath $U$}_f$ and
$\mbox{\protect\boldmath $S$}_f$ are not
uniquely determined, unless their unitarity is explicitely imposed.}
$\mbox{\boldmath $M$}_d'=\mbox{\boldmath $H$}_d
\mbox{\boldmath $U$}_d=\mbox{\boldmath $S$}_d^\dagger
\mbox{\boldmath $\cM$}_d \mbox{\boldmath $S$}_d
\mbox{\boldmath $U$}_d$, where
$\mbox{\boldmath $H$}_d\equiv
\sqrt{\mbox{\boldmath $M$}_d'\mbox{\boldmath $M$}_d^{'\dagger}}$
is an hermitian positive-definite matrix,
while $\mbox{\boldmath $U$}_d$ is unitary.
$\mbox{\boldmath $H$}_d$ can be diagonalized
by a unitary matrix $\mbox{\boldmath $S$}_d$; the resulting matrix
$\mbox{\boldmath $\cM$}_d$ is diagonal,
hermitian and positive definite.
Similarly,
one has
$\mbox{\boldmath $M$}_u'= \mbox{\boldmath $H$}_u
\mbox{\boldmath $U$}_u= \mbox{\boldmath $S$}_u^\dagger
\mbox{\boldmath $\cM$}_u \mbox{\boldmath $S$}_u
\mbox{\boldmath $U$}_u$ and
$\mbox{\boldmath $M$}_l'= \mbox{\boldmath $H$}_l
\mbox{\boldmath $U$}_l= \mbox{\boldmath $S$}_l^\dagger
\mbox{\boldmath $\cM$}_l \mbox{\boldmath $S$}_l
\mbox{\boldmath $U$}_l$.
In terms of the diagonal mass matrices,
$\mbox{\boldmath $\cM$}_d=\mbox{\rm diag}(m_d,m_s,m_b,\ldots)$,
$\mbox{\boldmath $\cM$}_u=\mbox{\rm diag}(m_u,m_c,m_t,\ldots)$,
$\mbox{\boldmath $\cM$}_l=\mbox{\rm diag}(m_e,m_\mu,m_\tau,\ldots)$,
the Yukawa Lagrangian takes the
simpler form
\bel{eq:N_Yuk_diag}
\cL_Y = - \left(1 + {H\over v}\right)\,\left\{
\overline{\mbox{\boldmath $d$}} \mbox{\boldmath $\cM$}_d
\mbox{\boldmath $d$} \, + \,
\overline{\mbox{\boldmath $u$}} \mbox{\boldmath $\cM$}_u
\mbox{\boldmath $u$} \, + \,
\overline{\mbox{\boldmath $l$}} \mbox{\boldmath $\cM$}_l
\mbox{\boldmath $l$} \right\} ,
\ee
where the mass eigenstates are defined by
\beqn\label{eq:S_matrices}
\mbox{\boldmath $d$}_L &\!\!\!\!\equiv &\!\!\!\!
\mbox{\boldmath $S$}_d\, \mbox{\boldmath $d$}'_L \, ,
\qquad\,\,\,\,\,\,\,\,\,
\mbox{\boldmath $u$}_L \equiv \mbox{\boldmath $S$}_u \,
\mbox{\boldmath $u$}'_L \, ,
\qquad\,\,\,\,\,\,\,\,\,
\mbox{\boldmath $l$}_L \equiv \mbox{\boldmath $S$}_l \,
\mbox{\boldmath $l$}'_L \, ,
\no\\
\mbox{\boldmath $d$}_R &\!\!\!\!\equiv &\!\!\!\!
\mbox{\boldmath $S$}_d \mbox{\boldmath $U$}_d\,
\mbox{\boldmath $d$}'_R \, , \qquad
\mbox{\boldmath $u$}_R \equiv \mbox{\boldmath $S$}_u
\mbox{\boldmath $U$}_u\, \mbox{\boldmath $u$}'_R \, , \qquad
\mbox{\boldmath $l$}_R \equiv \mbox{\boldmath $S$}_l
\mbox{\boldmath $U$}_l \, \mbox{\boldmath $l$}'_R \, .
\eeqn
Note, that the Higgs couplings are proportional to the
corresponding fermions masses.

Since, $\overline{\mbox{\boldmath $f$}}'_L \mbox{\boldmath $f$}'_L =
\overline{\mbox{\boldmath $f$}}_L \mbox{\boldmath $f$}_L$ and
$\overline{\mbox{\boldmath $f$}}'_R \mbox{\boldmath $f$}'_R =
\overline{\mbox{\boldmath $f$}}_R \mbox{\boldmath $f$}_R\,$
($f=d,u,l$), the form of the neutral-current part of the
$SU(2)_L\otimes U(1)_Y$ Lagrangian does not change when expressed
in terms of mass eigenstates. Therefore, there are no
flavour-changing neutral currents in the SM. This generalized GIM
mechanism is a consequence of treating all equal-charge fermions
on the same footing.

However, $\overline{\mbox{\boldmath $u$}}'_L \mbox{\boldmath $d$}'_L =
\overline{\mbox{\boldmath $u$}}_L \mbox{\boldmath $S$}_u
\mbox{\boldmath $S$}_d^\dagger
\mbox{\boldmath $d$}_L\equiv
\overline{\mbox{\boldmath $u$}}_L \mbox{\boldmath $V$}
\mbox{\boldmath $d$}_L$. In general,
$\mbox{\boldmath $S$}_u\not= \mbox{\boldmath $S$}_d$; thus
if one writes the weak eigenstates in terms of mass eigenstates,
a $N_G\times N_G$ unitary mixing matrix {\boldmath $V$},
called the Cabibbo--Kobayashi--Maskawa (CKM) matrix
\cite{cabibbo,KM:73}, appears in
the quark charged-current sector:
\bel{eq:cc_mixing}
\cL_{\mbox{\rms CC}}\, = \, {g\over 2\sqrt{2}}\,\left\{
W^\dagger_\mu\,\left[\sum_{ij}\,
\bar u_i\gamma^\mu(1-\gamma_5) \mbox{\boldmath $V$}_{ij} d_j
\, +\,\sum_l\, \bar\nu_l\gamma^\mu(1-\gamma_5) l
\right]\, + \, \mbox{\rm h.c.}\right\}\, .
\ee
The matrix {\boldmath $V$} couples any ``up-type'' quark with all
``down-type'' quarks.

Since neutrinos are massless,
we can always
redefine the neutrino flavours, in such a way as to eliminate
the analogous mixing in the lepton sector:
$\overline{\mbox{\boldmath $\nu$}}_L' \mbox{\boldmath $l$}'_L =
\overline{\mbox{\boldmath $\nu$}}_L' \mbox{\boldmath $S$}^\dagger_l
\mbox{\boldmath $l$}_L
\equiv
\overline{\mbox{\boldmath $\nu$}}_l \mbox{\boldmath $l$}_L$.
Thus, we have lepton-flavour conservation in the minimal SM
without right-handed neutrinos.

The fermion masses and the quark-mixing matrix {\boldmath $V$}
are all determined by the Yukawa couplings in Eq.~\eqn{eq:N_Yukawa}.
However, the Yukawas are not known; therefore we have a bunch of
arbitrary parameters.
A general $N_G\times N_G$ unitary matrix contains $N_G^2$ real
parameters [$N_G (N_G-1)/2$ moduli and $N_G (N_G+1)/2$ phases].
In the case of {\boldmath $V$}, many of these parameters are
irrelevant, because we can always choose arbitrary
quark phases.
Under the phase redefinitions $u_i\to \e^{i\phi_i} u_i$ and
$d_j\to\e^{i\theta_j} d_j$, the mixing matrix changes as
$\mbox{\boldmath $V$}_{ij}\to \mbox{\boldmath
$V$}_{ij}\e^{i(\theta_j-\phi_i)}$;
thus, $2 N_G-1$ phases are unobservable.
The number of physical free parameters in the quark-mixing matrix
gets then reduced to $(N_G-1)^2$:
$N_G(N_G-1)/2$ moduli and $(N_G-1)(N_G-2)/2$ phases.

In the simpler case of two generations, {\boldmath $V$}
is determined by a single parameter. One recovers then the
rotation Cabibbo matrix of Eq.~\eqn{eq:c_mixing}.
With $N_G=3$, the CKM matrix is described by 3 angles and 1 phase.
Different (but equivalent) representations can be found in the literature.
The Particle data Group \cite{pdg:94} advocates the use of the
following one as the ``standard'' CKM parametrization:
\be
\mbox{\boldmath $V$}\, = \, \left[
\begin{array}{ccc}
c_{12}c_{13}  &  s_{12} c_{13}  &  s_{13} e^{-i\delta_{13}} \\
-s_{12} c_{23}-c_{12}s_{23}s_{13}e^{i\delta_{13}} &
c_{12} c_{23}- s_{12} s_{23} s_{13} e^{i\delta_{13}} &
s_{23} c_{13}  \\
s_{12} s_{23}-c_{12}c_{23}s_{13}e^{i\delta_{13}} &
-c_{12} s_{23}- s_{12} c_{23} s_{13} e^{i\delta_{13}} &
c_{23} c_{13}
\ea
\right] .
\ee
Here $c_{ij} \equiv \cos{\theta_{ij}}$ and
$s_{ij} \equiv \sin{\theta_{ij}}$, with $i$ and $j$ being
``generation'' labels ($i,j=1,2,3$).
The real angles $\theta_{12}$, $\theta_{23}$ and $\theta_{13}$
can all be made to lie in the first quadrant, by an appropriate
redefinition of quark field phases; then,
$c_{ij}\geq 0$, $s_{ij}\geq 0$ and $0\leq \delta_{13}\leq 2\pi$.

Notice that $\delta_{13}$ is the only complex phase in the SM
Lagrangian. Therefore, it is the only possible source of CP-violation
phenomena. In fact, it was for this reason that the third generation
was assumed to exist \cite{KM:73},
before the discovery of the $b$ and the $\tau$.
With two generations, the SM could not explain the observed
CP-violation in the $K$ system.

\subsection{Standard Model parameters}
\label{subsec:parameters}

In the gauge  and scalar sectors, the SM Lagrangian contains only
4 parameters: $g$, $g'$, $\mu^2$ and $h$. We can trade these
parameters by $\alpha$, $\theta_W$, $M_W$ and $M_H$.
Alternatively, one can choose as free parameters
$\alpha$, $M_Z$, $G_F$ and $M_H$; this has the advantage of using the 3
most precise experimental determinations to fix the interaction.
In any case, one describes a lot of physics with only 4 inputs.

In the Yukawa sector, however, the situation is very different.
With $N_G=3$, we have 13 free parameters: 9 masses, 3 angles and 1 phase.
Clearly, this is not very satisfactory. The source of this proliferation
of parameters is the set of unknown Yukawa couplings in Eq.~\eqn{eq:N_Yukawa}.
The origin of masses and mixings, together with the reason
for the existing family replication, constitute at present
the main open problem in electroweak physics.

Thus, taking into account the QCD coupling constant $\alpha_s(M_Z^2)$,
the complete
$SU(3)_C \otimes SU(2)_L \otimes U(1)_Y$
SM Lagrangian is determined by 18 free parameters
(19 if one considers also a possible CP-violating $\bar\theta$ term
in the strong Lagrangian).


\setcounter{equation}{0}
\section{Tree-level Phenomenology}
\label{sec:tree}

It is convenient to take as inputs the well-measured quantities
\cite{pdg:94,lep:94}:
\beqn\label{eq:SM_inputs}
G_F & = & (1.16639 \pm 0.00002) \times 10^{-5}\,\mbox{\rm GeV}^{-2} ,
\no\\
\alpha^{-1} & = & 137.0359895\pm 0.0000061 ,
\\
M_Z & = & (91.1888\pm 0.0044)\,\mbox{\rm GeV} . \no
\eeqn
The relations
\bel{eq:A_def}
M_W^2 s_W^2  =  {\pi\alpha\over\sqrt{2} G_F}\equiv
A =  [(37.2802\pm 0.0003)\,\mbox{\rm GeV}]^2 ,
\qquad\qquad
s_W^2  =  1 - {M_W^2\over M_Z^2}\, ,
\ee
determine $s_W\equiv\sin{\theta_W}$ and $M_W$:
\bel{eq:M_W_det}
\!\!\!\!
M_W  =  {M_Z\over\sqrt{2}}\,
\left\{1 + \sqrt{1-{4A\over M_Z^2}}\right\}^{1/2}
 =  80.94\, \mbox{\rm GeV} , \qquad\quad
s_W^2  =  {1\over 2}\, \left\{1 - \sqrt{1-{4A\over M_Z^2}}\right\}
 =  0.2121 \, .
\ee
The predicted $M_W\, $is in
good agreement with the measured value, $M_W = 80.23\pm 0.18$ GeV.

At tree level, the decay widths of the weak gauge bosons can be
easily computed:
\beqn\label{eq:W_width}
&&\!\!\!\!\!\!\!\!\!\!\!
\Gamma\left[W^-\to \left(\ba \bar\nu_l l^- \\ \bar u_i d_j
\ea\right) \right]
 =  {G_F M_W^3\over 6\pi\sqrt{2}} \,
\left(\ba 1 \\ |V_{ij}|^2 N_C \ea\right)  = 0.2320\,
\left(\ba 1 \\ |V_{ij}|^2 N_C \ea\right)
\, \,\mbox{\rm GeV} , \quad
\\ \label{eq:Z_width}
&&\!\!\!\!\!\!\!\!
\Gamma\left[ Z\to \bar f f\right]  =
{G_F M_Z^3\over 6\pi\sqrt{2}} \, \left(|v_f|^2 + |a_f|^2\right)\,
N_f  =  0.3318\, \left(|v_f|^2 + |a_f|^2\right)\, N_f
\, \,\mbox{\rm GeV} ,
\eeqn
where $N_l=1$ and $N_q=N_C$.
Summing over all possible final fermion pairs, one predicts
the total widths
$\Gamma_W=2.09$ GeV and $\Gamma_Z=2.474$ GeV, in excellent agreement
with the experimental values \cite{pdg:94,lep:94}
$\Gamma_W=(2.08\pm 0.07)$ GeV
and $\Gamma_Z=(2.4974\pm 0.0038)$ GeV.

The universality of the $W$ couplings implies
\bel{eq:W_lep}
\mbox{\rm Br}(W^-\to\bar\nu_l l^-)\, = \,
{1\over 3 + 2 N_C}\, = \, 11.1 \% \, ,
\ee
where we have taken into account that the decay into the top quark
is kinematically forbidden,
$m_t=174\pm10{\,}^{+13}_{-12}$ GeV \cite{CDF:94}.
Similarly, the leptonic decay widths of the $Z$ are predicted
to be
\bel{eq:Z_lep}
\Gamma_l\,\equiv\,\Gamma(Z\to l^+l^-)\, = \,
84.85\,\, \mbox{\rm MeV} .
\ee
As shown in Table~\ref{tab:leptonic_modes},
the predictions \eqn{eq:W_lep}
and \eqn{eq:Z_lep} are in excellent agreement with the measured leptonic
widths. Moreover, the data confirms the universality of the
$W$ and $Z$ leptonic couplings at the 9\% and 0.4\% level, respectively.
If lepton universality is assumed, the experimental average of the
three leptonic modes gives\footnote{
$\Gamma_l$ refers to the partial decay width into a pair of massless
charged leptons.
}
$\mbox{\rm Br}(W^-\to\bar\nu_l l^-)=(10.76\pm 0.33)\% $ and
$\Gamma_l = 83.96\pm 0.18$ MeV \cite{pdg:94,lep:94}.

\begin{table}[htb]
\begin{center}
\caption{Measured \protect\cite{pdg:94,lep:94} values
of $\mbox{\rm Br}(W^-\to\bar\nu_l l^-)$, $\Gamma(Z\to l^+l^-)$
and the leptonic forward-backward asymmetries.
The average of the three leptonic modes is shown in the last column.
\label{tab:leptonic_modes}}
\vspace{0.2cm}
\begin{tabular}{|c||c|c|c||c|}
\hline
& $e$ & $\mu$ & $\tau$ & $l$
\\ \hline\hline
Br($W^-\to\bar\nu_l l^-$) \, (\%) & $10.8\pm 0.4$
& $10.6\pm 0.7$ & $10.8\pm 1.0$ & $10.76\pm 0.33$
\\
$\Gamma(Z\to l^+l^-)$ \, (MeV) & $83.85\pm 0.21$
& $83.95\pm 0.30$ & $84.26\pm 0.34$ & $83.96\pm 0.18$
\\
$\cA_{\mbox{\rms FB}}^{0,l}$ \, (\%) & $1.56\pm 0.34$
& $1.41\pm 0.21$ & $2.28\pm 0.26$ & $1.70\pm 0.16$
\\ \hline
\end{tabular}
\end{center}
\end{table}

Other useful quantities are the $Z$-decay width into invisible modes,
\bel{eq:Z_inv}
 {\Gamma_{\mbox{\rms inv}}\over\Gamma_l}\,\equiv\,
{N_\nu\,\Gamma(Z\to\bar\nu\nu)\over\Gamma_l}\, = \,
{N_\nu\over 2 (|v_l|^2 + |a_l|^2)} \, = \, 5.865 \, ,
\ee
which is usually normalized to the (charged) leptonic width,
and the ratio
\bel{eq:Z_had}
R_Z\,\equiv\, {\Gamma(Z\to\mbox{\rm hadrons})\over\Gamma_l}\, = \,
20.29 \, .
\ee
The comparison with the experimental values, shown  in
Table~\ref{tab:results}, is excellent.

\subsection{Fermion-pair production at the $Z$ peak}
\label{subsec:Zpeak}

Additional information can be obtained from the study of the
fermion-pair production process
\bel{eq:production}
e^+e^-\to\gamma,Z\to\bar f f \, .
\ee
For unpolarized $e^+$ and $e^-$ beams, the differential production
cross-section can be written, at lowest order, as
\bel{eq:dif_cross}
{d\sigma\over d\Omega}\, = \, {\alpha^2\over 8 s} \, N_f \,
         \left\{ A \, (1 + \cos^2{\theta}) \, + B\,  \cos{\theta}\,
     - \, h_f \left[ C \, (1 + \cos^2{\theta}) \, +\, D \cos{\theta}
         \right] \right\} ,
\ee
where $h_f$ ($=\pm1$) is the helicity of the produced fermion $f$,
and $\theta$ is the scattering angle between $e^-$ and $f$.
Here,
\beqn\label{eq:A}
A & = & 1 + 2 v_e v_f \mbox{\rm Re}(\chi)
 + \left(v_e^2 + a_e^2\right) \left(v_f^2 + a_f^2\right) |\chi|^2,
\\ \label{eq:B}
B & = & 4 a_e a_f \mbox{\rm Re}(\chi) + 8 v_e a_e v_f a_f  |\chi|^2  ,
\\ \label{eq:C}
C & = & 2 v_e a_f \mbox{\rm Re}(\chi) + 2 \left(v_e^2 + a_e^2\right) v_f
      a_f |\chi|^2 ,
\\ \label{eq:D}
D & = & 4 a_e v_f \mbox{\rm Re}(\chi) + 4 v_e a_e \left(v_f^2 +
      a_f^2\right) |\chi|^2  ,
\eeqn
and  $\chi$  contains the $Z$  propagator
\bel{eq:Z_propagator}
\chi \, = \, {G_F M_Z^2 \over 2 \sqrt{2} \pi \alpha }
     \,\, {s \over s - M_Z^2 + i s \Gamma_Z  / M_Z } \, .
\ee

The coefficients $A$, $B$, $C$ and $D$ can be experimentally determined,
by measuring the total cross-section, the forward-backward asymmetry,
the polarization asymmetry and the forward-backard polarization
asymmetry, respectively:
\beqn\label{eq:sigma}
\quad \sigma(s)
\,\,\,\,\,\,\,\,\,\, \,\,\,\,\,\, &
\!\!\!\!\!\!\!\!\!\!\! \!\!\!\!\!\!\!\!\!\!\!\!\!\!\!\!\!\!\!\! \!\! \! = \!
&\!\!\! \!\!\!\!\!\!\!\!\!\!\!\!\!\!
{4 \pi \alpha^2 \over 3 s } \, N_f \, A \, ,
\qquad\qquad\qquad\qquad
\cA_{\mbox{\rms FB}}(s)\equiv {N_F - N_B \over N_F + N_B}
     =  {3 \over 8} {B \over A}\,  ,
\\ \label{eq:A_pol}
&& \cA_{\mbox{\rms Pol}}(s)  \equiv
{\sigma^{(h_f =+1)}
- \sigma^{(h_f =-1)} \over \sigma^{(h_f =+1)} + \sigma^{(h_f = -1)}}
 =  - {C \over A} \, ,
\\ \label{eq:A_FB_pol}
\cA_{\mbox{\rms FB,Pol}}(s) &\!\!\! \equiv &\!\!\!
{N_F^{(h_f =+1)} -
N_F^{(h_f = -1)} - N_B^{(h_f =+1)} + N_B^{(h_f = -1)} \over
 N_F^{(h_f =+1)} + N_F^{(h_f = -1)} + N_B^{(h_f =+1)} + N_B^{(h_f = -1)}}
  =  -{3 \over 8} {D \over A}\, . \quad
\eeqn
Here, $N_F$ and $N_B$ denote the number of $f$'s
emerging in the forward and backward hemispheres,
respectively, with respect to the electron direction.
The measurement of the final fermion polarization can be done for $f=\tau$,
by measuring the distribution of the final $\tau$-decay products.

For $s = M_Z^2$,
the real part of the $Z$-propagator vanishes
and the photon exchange terms can be neglected
in comparison with the $Z$-exchange contributions
($\Gamma_Z^2 / M_Z^2 << 1$). Eqs. \eqn{eq:sigma} to \eqn{eq:A_FB_pol}
become then,
\beqn\label{eq:sigma_Z}
\sigma^{0,f}  \equiv  \sigma(M_Z^2)  =
 {12 \pi  \over M_Z^2 } \, {\Gamma_e \Gamma_f\over\Gamma_Z^2}\, ,
&& \qquad\qquad
\cA_{\mbox{\rms FB}}^{0,f}\equiv\cA_{FB}(M_Z^2) = {3 \over 4}
\cP_e \cP_f \, ,
\\ \label{eq:A_pol_Z}
\cA_{\mbox{\rms Pol}}^{0,f} \equiv
  \cA_{\mbox{\rms Pol}}(M_Z^2)  = \cP_f \, ,\quad
&& \qquad\qquad
\cA_{\mbox{\rms FB,Pol}}^{0,f} \equiv
\cA_{\mbox{\rms FB,Pol}}(M_Z^2)  =  {3 \over 4} \cP_e  \, ,
\eeqn
where $\Gamma_f$ is the $Z$ partial decay width to the $\bar f f$ final state,
and
\bel{eq:P_f}
\cP_f \, \equiv \, { - 2 v_f a_f \over v_f^2 + a_f^2}
\ee
is the average longitudinal polarization of the fermion $f$,
which only depends on the ratio of the vector and axial-vector couplings.
$\cP_f$ is a sensitive function of $\sin^2{\theta_W}$.

With polarized $e^+e^-$ beams, one can also study the ``left-right''
asymmetry between the cross-sections for initial left- and right-handed
electrons.
At the $Z$ peak, this asymmetry directly measures
the average initial lepton polarization, $\cP_e$,
without any need for final particle identification:
\bel{eq:A_LR}
\cA_{\mbox{\rms LR}}^0\,\equiv\, \cA_{\mbox{\rms LR}}(M_Z^2)
  \, = \, {\sigma_L(M_Z^2)
- \sigma_R(M_Z^2) \over \sigma_L(M_Z^2) + \sigma_R(M_Z^2)}
\, = \,  - \cP_e \,  .
\ee
%

\begin{table}[htb]
\begin{center}
\caption{Comparison between tree-level SM predictions and
experimental \protect\cite{pdg:94,lep:94,slc:94,MW:94} measurements.
The third column shows the effect
of including the main QED and QCD corrections.
The experimental value for $s_W^2$ refers to the
effective electroweak mixing angle in the charged-lepton sector,
defined in Eq.~\protect\eqn{eq:bar_s_W_l}.
\label{tab:results}}
\vspace{0.2cm}
\begin{tabular}{|c||c|c||c|}
\hline
Parameter & \multicolumn{2}{c||}{Theoretical prediction} &
Experimental
\\ \cline{2-3} & Tree-level & improved & value \\ \hline\hline
$M_W$ \,\, (GeV) & 80.94 & 79.95 & $80.23\pm 0.18$
\\
$s_W^2$ & 0.2121 & 0.2314 & $0.2320\pm 0.0004$
\\
$\Gamma_W$ \,\, (GeV) & 2.09 & 2.06 & $2.08\pm 0.07$
\\
$\Gamma_Z$ \,\, (GeV) & 2.474 & 2.486 & $2.4974\pm 0.0038$
\\
Br($W^-\to\bar\nu_l l^-$) \,\, (\%) & 11.1 & 10.8 & $10.76\pm 0.33$
\\
$\Gamma_l$ \,\, (MeV) & 84.85 & 83.41 & $83.96\pm 0.18$
\\
$\Gamma_{\mbox{\rms inv}}/\Gamma_l$ & 5.865 & 5.967 & $5.953\pm 0.046$
\\
$R_Z$ & 20.29 & 20.84 & $20.795\pm 0.040$
\\
$\sigma^0_{\mbox{\rms Had}}$ \,\, (nb) & 42.13 & 41.41 & $41.49\pm 0.12$
\\
$\cA_{\mbox{\rms FB}}^{0,l}$ & 0.0657 & 0.0165 & $0.0170\pm 0.0016$
\\
$\cA_{\mbox{\rms Pol}}^{0,\tau}$ & $-0.296$ & $-0.148$ & $-0.143\pm 0.010$
\\
${4\over 3}\cA^{0,\tau}_{\mbox{\rms FB,Pol}}$ & $-0.296$ & $-0.148$ &
$-0.135\pm 0.011$
\\
$-\cA_{\mbox{\rms LR}}^0$ & $-0.296$ & $-0.148$ & $-0.1637\pm 0.0075$
\\
$\cA_{\mbox{\rms FB}}^{0,b}$ & 0.210 & 0.104 & $0.0967\pm 0.0038$
\\
$\cA_{\mbox{\rms FB}}^{0,c}$ & 0.162 & 0.074 & $0.0760\pm 0.0091$
\\
$R_b$ & 0.219 & 0.220 & $0.2202\pm 0.0020$
\\
$R_c$ & 0.172 & 0.170 & $0.1583\pm 0.0098$
\\ \hline
\end{tabular}
\end{center}
\end{table}

Using the value of the weak mixing angle determined in Eq.~\eqn{eq:M_W_det},
one gets the predictions:
\bel{eq:sigma_had}
\!\!\!\! \sigma^0_{\mbox{\rms Had}} \equiv
\sum_q \, \sigma^{0,q} =
42.13 \, \mbox{\rm nb}  ,
\quad\quad
\cA_{\mbox{\rms FB}}^{0,l}  =  0.0657 \, ,
\quad\quad
\cA_{\mbox{\rms Pol}}^{0,l} =
{4\over 3}\, \cA_{\mbox{\rms FB,Pol}}^{0,l} =
- \cA_{\mbox{\rms LR}}^0=  -0.296 \, .
\ee

The comparison with the experimental measurements, given in
Table~\ref{tab:results}, is excellent for the total hadronic
cross-section; however, all leptonic asymmetries disagree with
the measured values by several standard deviations.
As shown in the table, the same happens with the
heavy-flavour asymmetries,
\bel{eq:A_b_c}
\cA_{\mbox{\rms FB}}^{0,b} \, = \, 0.210 \, , \qquad\qquad\qquad
\cA_{\mbox{\rms FB}}^{0,c} \, = \, 0.162\, ,
\ee
which compare very badly with the experimental measurements.
The partial hadronic widths,
\bel{eq:R_b_c}
R_b\,\equiv\, {\Gamma(Z\to\bar b b)\over \Gamma(Z\to\mbox{\rm hadrons})}
\, = \, 0.219 \, , \qquad\qquad
R_c\,\equiv\, {\Gamma(Z\to\bar c c)\over \Gamma(Z\to\mbox{\rm hadrons})}
\, = \, 0.172 \, ,
\ee
are again in good agreement with the data.
Clearly, the problem with the asymmetries is their high sensitivity
to the input value of $\sin^2{\theta_W}$. Therefore, they are an
extremely good window into higher-order electroweak corrections.

\subsection{Important QED and QCD corrections}
\label{subsec:QED_QCD_corr}

Before trying to analyze the relevance of higher-order electroweak
contributions, it is instructive to consider the numerical impact
of the well-known QED and QCD corrections.

The photon propagator gets vacuum polarization corrections, induced by
virtual fermion-antifermion pairs. The conservation of the
electromagnetic current, together with Lorentz invariance, imply that
the 1--loop vacuum-polarization amplitude takes the form
\bel{eq:vac_pol}
-i \,\Pi^{\mu\nu}(q^2) \, =\, -i \left( -g^{\mu\nu} q^2 + q^\mu q^\nu\right)
\, \Pi_\gamma(q^2)\, ,
\ee
with $\Pi_\gamma(q^2)$ a $\cO(\alpha)$ scalar function satisfying
$\Pi_\gamma(0)=0$.
Remembering the form of the tree-level photon propagator,
$-i g^{\alpha\beta}/q^2$, it is straightforward to sum the effect
of an infinite number of 1--loop vacuum-polarization insertions.
The correction induced in the $e^+e^-\to\gamma\to \bar f f$ process
amounts to the change:
\bel{eq:alpha_running}
{\alpha\over s}\,\Longrightarrow\,
{\alpha\over s}\,\left\{ 1 - \Pi_\gamma(s) + \Pi_\gamma^2(s) -
\Pi_\gamma^3(s) + \ldots\right\}
\, = \, {1\over s}\, {\alpha\over 1 + \Pi_\gamma(s)}\,\approx\,
{\alpha(s)\over s}\, .
\ee
Thus, we can take into account this kind of QED loop corrections, by
making a redefinition of the QED coupling. But now, $\alpha(s)$ is a function
of the energy scale, i.e. the effective coupling ``runs'' with the
energy; $\alpha(s)$ is called the QED running coupling.
The fine structure constant in Eq.~\eqn{eq:SM_inputs} is measured
at very low energies; it corresponds to $\alpha(m_e^2)$.
However, at the $Z$ peak, we should rather use $\alpha(M_Z^2)$.
The long running from $m_e$ to $M_Z$ gives rise to a sizeable QED
correction \cite{BU:89}:
\bel{eq:QED_corr}
\alpha\equiv\alpha(m_e^2)\,\Longrightarrow\,
\alpha(M_Z^2)\equiv{\alpha\over 1-\Delta\alpha} = 1.064\,\alpha\, .
\ee

The running effect generates an important change in Eq.~\eqn{eq:A_def}.
Since $G_F$ is measured at low energies, while $M_W$ is a
high-energy parameter, the relation between both quantities is clearly
modified by vacuum-polarization contributions:
\bel{eq:barA_def}
M_W^2 s_W^2 \, = \, {\pi\alpha(M_Z^2)\over\sqrt{2} G_F}
\, = \, {A\over 1 -\Delta\alpha}
\,\equiv\,
\bar A\, = \, [38.455\,\mbox{\rm GeV}]^2 ,
\ee
Changing $A$ by $\bar A$ in Eqs.~\eqn{eq:M_W_det},
one gets the corrected predictions:
\bel{eq:new_pred}
M_W\, = \, 79.95\,\mbox{\rm GeV} , \qquad\qquad\qquad
s^2_W \, = \, 0.2314 \, .
\ee
The value of $M_W$ is now in better agreement with the
experimental determination.

So far, we have treated quarks and leptons on an equal footing.
However, quarks are strong-interacting particles.
The gluonic corrections to the decays
$Z\to\bar q q$ and $W^-\to\bar u_i d_j$ can be directly incorporated
into the formulae given before, by taking an ``effective'' number of
colours:
\bel{N_C_eff}
N_C \, \Longrightarrow \,
N_C\,\left\{ 1 + {\alpha_s\over\pi} + \ldots\right\}\,
\approx\, 3.115 \, ,
\ee
where we have used $\alpha_s(M_Z^2)\approx 0.12\, $.
Note that the strong coupling also ``runs''; one should then use
the value of $\alpha_s$ at $s=M_Z^2$.

The third column in Table~\ref{tab:results} shows the numerical impact
of these  QED and QCD corrections. In all cases, the comparison with
the data gets improved. However, it is in the asymmetries where the
effect gets more spectacular. Owing to the high sensitivity to $s^2_W$,
the small change in the value of the weak mixing angle generates
a huge difference of about a factor of 2 in the predicted
asymmetries.
The agreement with the experimental values is now very good.

\setcounter{equation}{0}
\section{Higher-Order Electroweak Corrections}
\label{sec:nc-loop}

We can distinguish five types of loop corrections:
\begin{enumerate}
\item {\bf QED:} Initial- and final-state photon radiation is by far the
most important numerical correction. One has in addition the contributions
coming from photon exchange between the fermionic lines.
All these corrections are to a large extent dependent on the detector and
the experimental cuts, because of the infra-red problems associated with
massless photons (one needs to define, for instance, the minimun photon-energy
which can be detected). Therefore, these effects are usually estimated with
Monte Carlo programs and subtracted from the data.
Notice that in the decay $\mu^-\to e^- \bar\nu_e \nu_\mu$,
the QED corrections are already partly included in the definition of $G_F$
[see Eq.~\eqn{eq:mu_lifetime}];
thus, one should take care of subtracting those
corrections already incorporated in the old $V-A$ calculation.
\item {\bf Oblique:} The gauge-boson self-energies, induced by vacuum
polarization diagrams. We have already seen the important role of the
photon self-energy in the previous section. In the case of the $W^\pm$ and
the $Z$, these corrections are very interesting because they are sensitive
to heavy particles (such as the top) running along the loop
\cite{VE:77}.
In addition, these contributions are ``universal'' (process independent).
\item {\bf Vertex:} Corrections to the different couplings.
They are ``non-universal'' and usually smaller than the oblique contributions.
There is one interesting exception, the $Z \bar bb$ vertex, which is sensitive
to the top quark mass \cite{BPS:88}.
\item {\bf Box:} Diagrams with two gauge-boson exchanges. At the $Z$ peak,
they give a very small contribution, because they are non resonant (they
do not have an on-shell $Z$ propagator). However, the box correction to the
decay $\mu^-\to e^- \bar\nu_e \nu_\mu$ is not negligible.
\item {\bf Higgs:} The exchange of a Higgs particle between two fermionic
lines.
This correction is usually irrelevant, because the amplitude is suppressed
by the product of the two fermionic masses.
\end{enumerate}

\subsection{The weak mixing angle}
\label{subsec:sinTheta}

At tree level, there are three different places where $\sin^2{\theta_W}$
appears:
\bi
\item In the charged-current sector, regulates the relation between $M_W$
and $G_F$ [Eq.~\eqn{eq:barA_def}].
\item The neutral couplings in Eq.~\eqn{eq:Z_Lagrangian}.
 Thus, it shows up in the vector couplings $v_f$ and in
the ratio of neutral to charged current interactions,
\bel{eq:rho}
\left( {\sigma_{\mbox{\rms NC}}(\nu f) \over \sigma_{\mbox{\rms CC}}(\nu f)}
\right)^{1/2} \,\sim\, \rho \, = \,\rho_0\,\equiv\,
{M_W^2\over M_Z^2 \cos^2{\theta_W}} \, .
\ee
\item In the SSB mechanism, which predicts $\rho_0=1$.
\ei

Quantum loops generate different corrections in the three sectors;
thus, one needs to specify how the weak mixing angle is defined.
The so-called ``on-shell'' scheme adopts the definition \cite{SI:80}:
\bel{eq:s_W_def}
s_W^2\,\equiv\, 1 - {M_W^2/ M_Z^2} \, .
\ee
The corrections to the different couplings are then parametrized as
\beqn\label{eq:D_r}
M_W^2 s_W^2 & = & M_Z^2 s_W^2 c_W^2 \, = \, {A\over 1 -\Delta r}\, ,
\\ \label{eq:D_rho}
\rho & = &{\rho_0\over 1 -\Delta\rho}\, =\, {1\over 1 -\Delta\rho} \, ,
\\ \label{eq:k_f}
v_f & = & T_3^f\, \left( 1 - 4 |Q_f| K_f s^2_W\right) \, .
\eeqn

\subsection{Oblique corrections}
\label{subsec:oblique}

Quantum corrections offer the possibility to be sensitive to
heavy particles, which cannot be kinematically accessed,
through their virtual loop effects.

In QED, the vacuum polarization contribution of a heavy fermion pair,
$\gamma\to \bar f f \to \gamma$, is suppressed by inverse powers of
the fermion mass,
\bel{eq:dec}
\Pi_\gamma(s) \, = \, {\alpha\over 15\pi}\, {s\over m_f^2} \, + \,
\cO\left( {s^2\over m_f^4}\right) , \qquad\qquad (s<<m_f^2) .
\ee
At low energies, the information on the heavy fermions is then lost.
This ``decoupling'' of the heavy fields happens in theories
like QED and QCD,
with only vector couplings and an exact gauge symmetry \cite{AC:75},
where the effects generated by the heavy particles can always be
reabsorbed into a redefinition of the low-energy parameters.

The SM involves, however, a broken chiral gauge symmetry. This
has the very interesting implication of avoiding the decoupling
theorem \cite{AC:75}. The vacuum polarization contributions
modify the $W^\pm$ and $Z$ propagators:
\bel{eq:W_Z_prop}
{-g^{\mu\nu} + \cdots\over s - M^2_{Z/W}}\,\,\Longrightarrow\,\,
{-g^{\mu\nu} + \cdots\over s - M^2_{Z/W} + \Sigma_{Z/W}(s)} .
\ee
The self-energies $\Sigma_{Z/W}(s)$ induced by a heavy top,
i.e. $W^-\to \bar t b \to W^-$ and $Z\to\bar t t \to Z$,
generate contributions to $\Delta r$ and $\Delta\rho$,
\bel{eq:d_r_top}
\Delta r  =  {\Sigma_W(0)\over M^2_W} \sim
 - {c^2_W\over s^2_W}\, \Delta\bar\rho
 +  \ldots ,
\qquad\qquad
\Delta\rho  =  {\Sigma_Z(0)\over M^2_Z} -
{\Sigma_W(0)\over M^2_W}\sim\Delta\bar\rho
 +  \ldots ,
\ee
which increase quadratically with the top mass \cite{VE:77},
\bel{eq:d_rho_bar}
\Delta\bar\rho \,\equiv\,
{\alpha N_C\over 16\pi s^2_W c^2_W}\, {m_t^2\over M_Z^2} .
\ee
Therefore, a heavy top does not
decouple.
Taking $m_t = 174$ GeV, the leading quadratic correction
$\Delta\bar\rho$
amounts to
a $-3\% $ (0.9\%)  contribution to $\Delta r$ ($\Delta\rho$).

The quadratic mass contribution originates in the strong breaking
of weak isospin generated by the top and bottom quark masses, i.e.
the effect is actually proportional to $m_t^2-m_b^2$.
There is, however, a smaller non-decoupling contribution even
for a degenerate heavy fermion doublet $F\equiv(U,D)$, with
$M_U = M_D >> M_Z$. The virtual production of those hypothetical
heavy particles would generate the correction:
$\Delta r \sim \alpha N_C^F/ (12\pi s^2_W) \sim
2.5\times 10^{-3}\, (N_C^F/ 3)$;
$\,\Delta\rho = 0$.

Owing to an accidental $SU(2)_R$ symmetry of the scalar sector
(the so-called custodial symmetry), the
virtual production of Higgs particles does not generate any
$M_H^2$ dependence at one loop (Veltman screening \cite{VE:77}).
The dependence on the Higgs mass is only logarithmic:
\bel{eq:heavy_Higgs}
\Delta r \,\sim\, - {c^2_W\over s^2_W}\,\Delta\rho\,\sim\,
{\alpha\over 16\pi s^2_W}\, {11\over 3}\,\left[
\ln{\left( {M_H^2\over M_W^2}\right) } - {5\over 6}\right] ,
\qquad\qquad (M_H^2>>M_W^2) .
\ee
The numerical size of the correction is $-0.0041$ (0.0098) for
$M_H=50$ (1000) GeV.

Therefore, within the SM, the oblique corrections contain a strong
dependence on $m_t$     
and a much smaller one on $M_H$.
In addition, they are sensitive to all kinds of heavy new physics;
i.e. heavy new particles which cannot be produced at present
energies, but which could generate virtual contributions to
$\Delta r$ and $\Delta\rho$.
Taking all SM electroweak contributions into account:
\bel{eq:oblique_con}
\Delta\alpha  =  0.060 \, , \qquad\quad
\Delta r  =  \Delta\alpha - {c_W^2\over s_W^2}\,\Delta\bar\rho
+\ldots \, ,
\qquad\quad \Delta\rho  =  \Delta\bar\rho  +  \ldots
\ee
The appearence of $\Delta\alpha$ in the $W$  (and $Z$) contribution
should not be a surprise, since the vector-vector self-energy diagram
is basically the same as in QED.

In addition, the self-energy mixing between the $Z$ and the $\gamma$,
$Z\to\bar f f\to\gamma$, produces a contribution to the factor $K_f$ in the
$Z\bar f f$ vector coupling $v_f$.
Although $K_f$ is a non-universal vertex correction, this particular
``oblique'' contribution does not depend on the fermion $f$.
It has become usual to include this effect by defining an
``effective'' $\bar s^2_W$,
\bel{eq:s_W_eff}
\bar s^2_W \,\equiv\, s^2_W - c_W s_W \,\mbox{\rm Re}\left(
{\Pi_{\gamma Z}(s)\over 1 +\Pi_\gamma(s)}\right) \, = \,
s_W^2 \, + \, c_W^2\,\Delta\bar\rho\, + \, \ldots \, ,
\ee
such that
$v_f = T_3^f\, \left( 1 - 4 |Q_f| \bar s^2_W\right)$.

Note that
$s_W^2 (1-\Delta r) \approx \bar s_W^2 (1-\Delta\alpha)$ and
$\bar c_W^2\approx c_W^2 (1-\Delta\bar\rho)$.
Therefore, to a very good approximation,
the relation \eqn{eq:D_r} can be written as
\bel{eq:r_improved}
\rho M_Z^2 \bar c_W^2 \bar s_W^2 \,\approx\,
{A\over 1 -\Delta\alpha} \, ,
\ee
where the hard $m_t^2$ dependence has nearly disappeared
(it is only present in $\rho$, where the effect is about $1\% $).
We can then conclude that the tree-level determinations of
$M_W^2$ and the weak mixing angle are in fact very accurate,
provided one incorporates the QED correction $\Delta\alpha$ and
the effective $\bar s_W^2$ is used.
This explains the excellent results we have obtained in
Sect. \ref{subsec:QED_QCD_corr}.

\subsection{Improved Born Approximation}
\label{subsec:Born}

It is possible to understand nearly all electroweak effects at the
$Z$ peak, by using the simplified (tree-level like) formulae:
\beqn\label{eq:IBA_decay}
&&\Gamma(Z\to\bar f f) \approx  {G_F M_Z^3\over 6\pi\sqrt{2}}\,
\rho\, N_f \, \left( 1 + v_f^2 \right) ,
\\ \label{eq:IBA_prod}
T(e^+e^-\to\bar f f) \approx &&\!\!\!\!\!\! \!
{4\pi\alpha(M_Z^2)\over s}
\, Q_e  J^\mu_{\mbox{\rms em}}(e)\,  Q_f J_\mu^{\mbox{\rms em}}(f)
\, +\, \sqrt{2} G_F M_Z^2 \rho\,
 {J^\mu_Z(e) \, J_\mu^Z(f)\over s-M_Z^2 + i s \Gamma_Z/M_Z} \, ,
\qquad\quad
\eeqn
with
\bel{eq:IBA_nc}
\!\! v_f\approx T_3^f\, \left( 1 - 4 |Q_f| \bar s^2_W\right) ,
\,\quad
a_f \approx T_3^f , \quad\,
\Delta\rho \approx \Delta\bar\rho  = {3 G_F m_t^2\over 8\pi^2\sqrt{2}}
= 0.0096 \, \left({m_t\over 175\, \mbox{\rm GeV}}\right)^2 ,
\ee
and
\bel{eq:s_bar_eff}
\bar s_W^2\, = \, {1\over 2}\,\left\{ 1 -\sqrt{1 -
{4 A\over \rho M_Z^2 (1-\Delta\alpha)}}\right\} .
\ee

This  ``Improved Born Approximation'' works with a precision better
than 1\% (compared with the exact loop results), except for the
$Z\bar b b$ vertex that we will discuss next.

\subsection{The $Z\to\bar b b$ vertex}
\label{subsec:Zbb}

The $Z\bar f f$ vertex gets 1--loop corrections where a virtual
$W^\pm$ is exchanged between the two fermionic legs. Since, the $W^\pm$
coupling changes the fermion flavour, the decays
$Z\to \bar d d, \bar s s, \bar b b$ get contributions with a top quark
in the internal fermionic lines,
i.e. $Z\to\bar t t\to \bar d_i d_i$. Notice that this mechanism can also
induce the flavour-changing neutral-current decays
$Z\to \bar d_i d_j$ with $i\not=j$.
These amplitudes are suppressed by the small CKM mixing factors
$|\bV^{\phantom{*}}_{\!\! tj}\bV^*_{\!\! ti}|^2$.
However, for the $Z\to\bar b b$ vertex, there is no suppression
because $|\bV_{\!\! tb}|\approx 1$ (see Sect.~\ref{sec:cc-quarks}).

The explicit calculation
\cite{BPS:88,ABR:86} shows the presence of hard $m_t^2$ corrections to
the $Z\to\bar b b$ vertex. This effect can be easily understood
\cite{BPS:88}
in non-unitary gauges where the unphysical charged scalar $\phi^{(\pm)}$
is present. The Yukawa couplings of the charged scalar to fermions
are proportional to the fermion masses; therefore, the exchange
of a virtual $\phi^{(\pm)}$ gives rise to a $m_t^2$ factor.
In the unitary gauge, the charged scalar has been ``eaten'' by the
$W^\pm$ field; thus, the effect comes now from the exchange of a
longitudinal $W^\pm$,
with terms proportional to $q^\mu q^\nu$ in the propagator that
generate fermion masses.

Since the $W^\pm$ couples only to left-handed fermions, the induced
correction is the same for the vector
and axial-vector $Z\bar b b$ couplings \cite{BPS:88}:
\beqn\label{eq:delta_b}
\delta v_b \, = \, \delta a_b
& = & -{\alpha\over 8\pi s_W^2}\,\left\{
{m_t^2\over M_W^2} + \left( {8\over 3} + {1\over 6 c_W^2}\right)\,
\ln{\left({m_t^2\over M_W^2}\right)} \, + \, \ldots \right\} ,
\no \\
\delta_b &\equiv& {\delta\Gamma(Z\to\bar b b)\over \Gamma(Z\to\bar b b)}
\, \approx \, -5\times 10^{-3}\, \left( {m_t^2\over M_Z^2} - {2\over 5}
\right) .
\eeqn
The second line is just a fit to the exact numerical result, which
works better than 1\% in the range of $m_t$ between 90 and
230 GeV.

The ``non-decoupling'' present in the
$Z\bar b b$ vertex is quite different from the one happening in
the boson self-energies.
The vertex correction does not have any dependence with the
Higgs mass. Moreover,
while any kind of new heavy particle
coupling to the gauge bosons would contribute to the $W$ and $Z$
self-energies, the possible new physics contributions to the
$Z\bar b b$ vertex are much more restricted and, in any case,
different.
Therefore, an independent experimental test of the two effects
would be very valuable in order to disentangle the possible
new physics contributions from the SM corrections.
In addition, since the ``non-decoupling'' vertex effect is related
to $W_L$-exchange, it is sensitive to the SSB
mechanism.

Theoretically, the cleanest way to separate the vertex
correction $\delta_b$ would be through the ratio \cite{BPS:88}
$\Gamma(Z\to\bar b b)/\Gamma(Z\to\bar s s)\approx 0.9949 (1+\delta_b)$.
Except for the small kinematical correction due to the different
final masses, all other corrections cancel.
Unfortunately, it is quite difficult to make an accurate selection
of $Z\to\bar s s$ events. Therefore, the ratio $R_b$ is usually
used in the experimental analysis.
Although there is some cancellation between the $\delta_b$ contributions
to $Z\to\bar bb$ and $Z\to\mbox{\rm hadrons}$, the sensitivity of $R_b$
to the vertex correction is still quite good, while the main
QCD corrections cancel \cite{BPS:88}.
Another possibility would be to measure the ratio of $Z\to\bar bb$
events to $Z\to\bar u u +\bar d d + \bar s s + \bar c c$
\cite{BCS:93}, where only the numerator gets a $\delta_b$
contribution; this would require a very good efficiency for selecting
$\bar bb $ events.

\subsection{SM electroweak fit}
\label{subsec:EWfit}

The partial widths of the $Z$ into leptons, the leptonic forward-backward
asymmetries, the $\tau$ polarization $\cA_{\mbox{\rms Pol}}^{0,\tau}$
and the $\tau$ polarization asymmetry $\cA_{\mbox{\rms FB,Pol}}^{0,\tau}$,
can all be combined to determine effective vector and
axial-vector couplings of the three charged leptons.
The asymmetries determine the ratio $v_l/a_l$, while the sum
$(v_l^2 + a_l^2)$ is derived from
\bel{eq:Z_l_QED}
\Gamma_l \, = \,
{G_F M_Z^3\over 6\pi\sqrt{2}} \, (|v_f|^2 + |a_f|^2)\,
\left(1 + \delta_l^{\mbox{\rms QED}}\right) ,
\ee
where $\delta_l^{\mbox{\rms QED}}=3\alpha/(4\pi)$
accounts for the final-state photonic corrections.
The signs of $v_l$ and $a_l$ are fixed by requiring $a_e<0$.

\begin{table}[b] 
\begin{center}
\caption{Measured \protect\cite{lep:94}
effective vector and axial-vector couplings
of the charged leptons. The last column gives the averaged results,
assuming lepton universality.
\label{tab:nc_measured}}
\vspace{0.2cm}
\begin{tabular}{|c||c|c|c||c|}
\hline
& $e$ & $\mu$ & $\tau$ & $l$
\\ \hline\hline
$v_l$ & $-0.0370\pm 0.0021$ &
$-0.0308\pm 0.0051$ & $-0.0386\pm 0.0023$ & $-0.0366\pm 0.0013$
\\
$a_l$ & $-0.50093\pm 0.00064$ & $-0.50164\pm 0.00096$ & $-0.5026\pm 0.0010$ &
$-0.50128\pm 0.00054$
\\ \hline
\end{tabular}
\end{center}
\end{table}

\begin{figure}[htb]
\centerline{\mbox{\epsfysize=10.0cm\epsffile{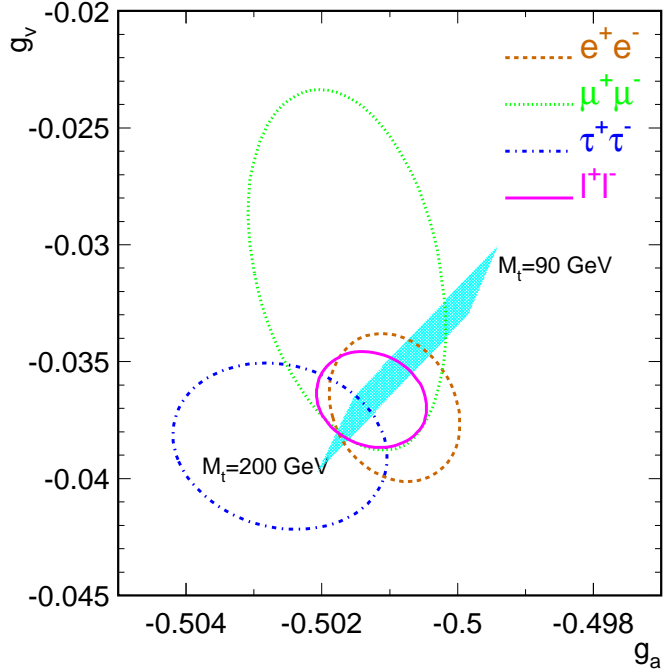}}}
\vspace{-1.0cm}
\caption{68\% probability contours in the $a_l$-$v_l$ plane. The solid contour
assumes
lepton universality. The shaded band represents
the SM prediction. (Taken from Ref. \protect\cite{lep:94})}
\label{fig:gagv}
\end{figure}

Table~\ref{tab:nc_measured} gives the averaged results obtained from the
present
LEP data \cite{lep:94}.
The 68\% probability contours in the $a_l$-$v_l$ plane are shown in
Fig.~\ref{fig:gagv}.
The measured ratios,
\bel{eq:nc_ratios}
\!\! {v_\mu\over v_e}=0.83\pm 0.16\, , \quad
{v_\tau\over v_e}=1.044\pm 0.091\, , \quad
{a_\mu\over a_e}=1.0014\pm 0.0021\, , \quad
{a_\tau\over a_e}=1.0034\pm 0.0023\, ,
\ee
provide a test of charged-lepton universality in the neutral-current sector.

The neutrino coupling can also be determined from the invisible $Z$-decay
width, by assuming three identical neutrino generations
with left-handed couplings, and fixing the sign from neutrino scattering
data \cite{charmII:94}.
The resulting experimental value \cite{lep:94},
\bel{eq:nu_coupling}
v_\nu\, = \, a_\nu \, = \, -0.5011\pm 0.0018 \, ,
\ee
is in perfect agreement with the SM.
Alternatively, one can use the SM prediction for $\Gamma_{\mbox{\rms inv}}$
to get a determination of the number of (light) neutrino flavours,
$N_\nu = 2.988\pm 0.023$ \cite{lep:94}.

The measured leptonic asymmetries can be used to obtain the
effective electroweak mixing angle in the charged-lepton sector \cite{lep:94}:
\bel{eq:bar_s_W_l}
\sin^2{\theta^{\mbox{\rms lept}}_{\mbox{\rms eff}}} \, \equiv \,
{1\over 4} \,  \left( 1 - {v_l\over a_l}\right) \, = \,
0.2317\pm 0.0007 \, .
\ee
Since the hadronic asymmetries have a reduced sensitivity
to corrections particular to the hadronic vertex,
$ \sin^2{\theta^{\mbox{\rms lept}}_{\mbox{\rms eff}}}$
can also be extracted from the quark forward-backward asymmetries.
The measured values of $\cA_{\mbox{\rms FB}}^{0,b}$
and $\cA_{\mbox{\rms FB}}^{0,c}$ imply
$\sin^2{\theta^{\mbox{\rms lept}}_{\mbox{\rms eff}}} =
0.2325\pm 0.0006$, while the total hadronic charge asymmetry
$\langle Q_{\mbox{\rms FB}}\rangle$ gives $0.2320\pm 0.0016$
\cite{lep:94}. The three LEP values are in good agreement,
giving and average of $0.2321\pm 0.0004$.

A slightly smaller value,
\bel{eq:s_eff_SLD}
\sin^2{\theta^{\mbox{\rms lept}}_{\mbox{\rms eff}}} \, =\,
0.2294\pm 0.0010 \, ,
\ee
is obtained from the measurement of $\cA_{\mbox{\rms LR}}$,
performed at SLC \cite{slc:94}.
Making a global average with the LEP determination, one gets
$\sin^2{\theta^{\mbox{\rms lept}}_{\mbox{\rms eff}}} =
0.2317\pm 0.0004$ with a $\chi^2/\mbox{\rm d.o.f.} = 9.0/6$.

Including the full SM predictions at the 1--loop level,
the $Z$ measurements can be used to obtain information on
the SM parameters.
Figs. \ref{fig:lep_results1} and \ref{fig:lep_results2},
taken from Ref.~\cite{lep:94},
compare different
LEP measurements with the corresponding SM predictions
as a function of $m_t$.
The parameters $M_H$ and $\alpha_s(M_Z^2)$ have been taken in the ranges
60 GeV $< M_H <$ 1000 GeV
and $\alpha_s(M_Z^2)=0.123\pm 0.006$.
For the comparison of $R_b$ with the SM the value of $R_c$ has been fixed
to the SM prediction.
The overall agreement is very good; moreover, the different observables
determine a similar range of $m_t$.
Only the ratios $R_b$ and $R_c$ seem to be slightly off.

\begin{figure}[t]  
\centerline{\mbox{\epsfysize=15.0cm\epsffile{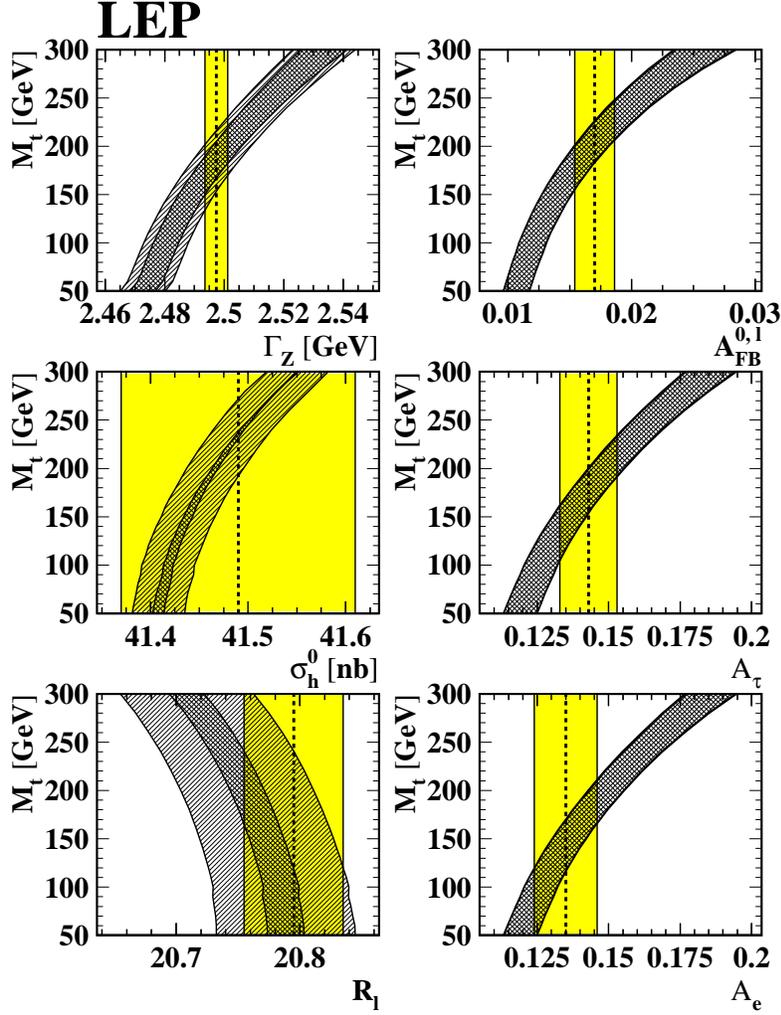}}}
\vspace{-1.5cm}
\caption{Comparison of LEP measurements with the SM prediction as a
function of $m_t$.
The cross-hatched area shows the variation of the SM prediction with
$M_H$ spanning the interval 60 GeV $< M_H <$ 1000 GeV and the singly-hatched
area corresponds to a variation of $\alpha_s(M_Z^2)$ within the interval
$\alpha_s(M_Z^2)=0.123\pm 0.006$. The total width of the band corresponds to
the linear sum of both uncertainties.
The experimental errors are indicated as
vertical bands.
(Taken from Ref.~\protect\cite{lep:94}).}
\label{fig:lep_results1}
\end{figure}


\begin{figure}[htb]
\centerline{\mbox{\epsfysize=15.0cm\epsffile{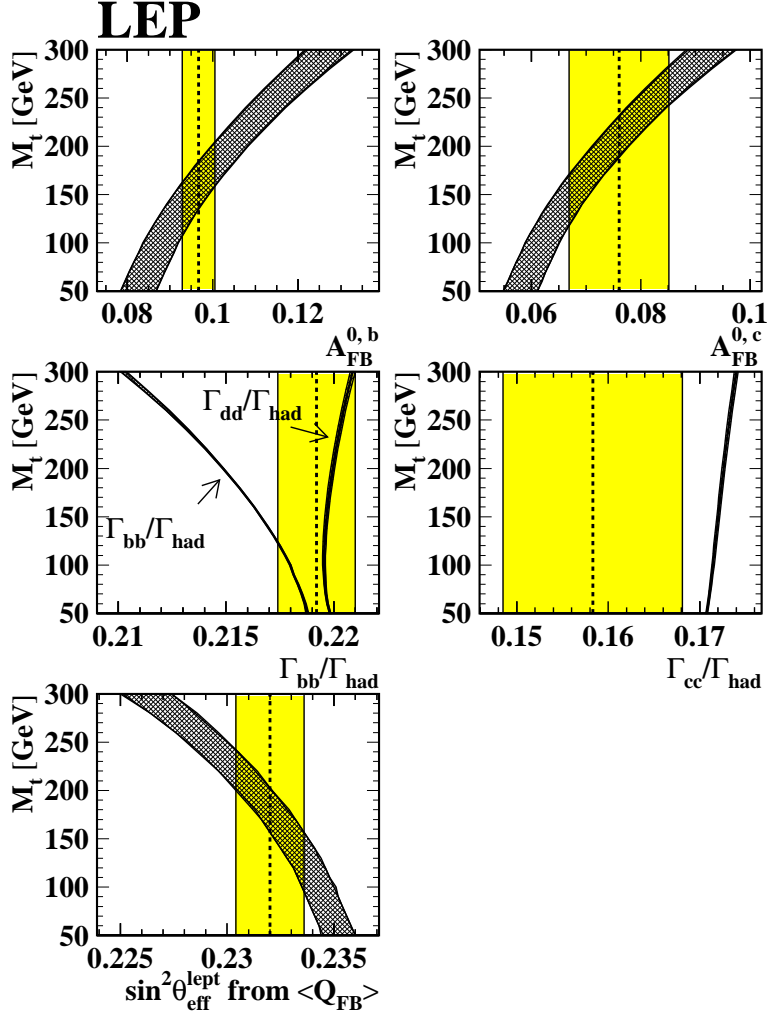}}}
\vspace{-1.35cm}
\caption{Same as Fig. \protect\ref{fig:lep_results1}.}
\label{fig:lep_results2}
\end{figure}

Table~\ref{tab:EW_fit} shows the constraints obtained on
$m_t$ and $\alpha_s(M_Z^2)$, from a global fit to the 
electroweak data \cite{lep:94}.
The fitted value of the top mass is in perfect agreement with the
CDF measurement, $m_t= 174\pm 10 {\,}^{+13}_{-12}$ GeV \cite{CDF:94}.
Moreover, the extracted value of the strong coupling agrees very well
with direct determinations from event
shape measurements,  $\alpha_s(M_Z^2) = 0.123\pm 0.006$ \cite{sorrento}, or
$\tau$ decay, $\alpha_s(M_Z^2) = 0.119\pm 0.004$ \cite{QCD:94}.

Taking the central values for $m_t$ and $\alpha_s(M_Z^2)$,
the SM fit implies $R_b= 0.2158$ and $R_c=0.172$.
The discrepancy with the measured asymmetries
$R_b=0.2202\pm 0.0020$ and $R_c=0.1583\pm 0.0098$ is at the $2.2\sigma$
and $-1.4\sigma$ level, respectively.
The agreement between $R_b$ and its SM prediction improves to $1.9\sigma$
if $R_c$ is fixed to the SM value of $R_c=0.171$ (there is a strong
correlation between both hadronic asymmetries). In this case,
one obtains $R_b=0.2192\pm 0.0018$ \cite{lep:94}.
Although it is still premature to extract any conclusion from this
small (and statistically not significant) discrepancy, it is
certainly an important thing to keep in mind, in view of the
special role played by the $Z\bar bb$ vertex.

Since the main $M_H$ dependence is only logarithmic, it is not
possible to extract information on $M_H$, without a direct
measurement of $m_t$. If the CDF value of $m_t$ is used as an
additional constraint on the electroweak fit, the observed
$\Delta\chi^2\equiv\chi^2-\chi^2_{\mbox{\rms min}}$ curve exhibits
a minimum for low values of $M_H$. However, at 95\%
CL the entire range of $M_H$ up to
1000 GeV is allowed.

\begin{table}[hbt]
\begin{center}
\caption{Values of $m_t$ and $\alpha_s(M_Z^2)$ obtained from
a global fit to present electroweak precision data
\protect\cite{lep:94}.
No external constraint on $\alpha_s(M_Z^2)$ has been imposed.
The central values and the first errors quoted refer to $M_H=300$ GeV.
The second errors correspond to the variation of the central value
in the interval 60 GeV $< M_H <$ 1000 GeV.
The bottom part of the table lists derived results for
the effective leptonic electroweak mixing angle,
$1-M_W^2/M_Z^2$ and $M_W$.
\label{tab:EW_fit}}
\vspace{0.2cm}
\begin{tabular}{|c||c|c|c|}
\hline
& LEP & LEP & LEP
\\
& only & + $p\bar p$ and $\nu N$ data & + $p\bar p$ and $\nu N$ data +
$\cA_{\mbox{\rms LR}}$
\\ \hline\hline
$m_t$ \,\, (GeV) & $173 {\,}^{+12}_{-13}{\,}^{+18}_{-20}$ &
$171{\,}^{+11}_{-12}{\,}^{+18}_{-19}$ & $178{\,}^{+11}_{-11}{\,}^{+18}_{-19}$
\\
$\alpha_s(M_Z^2)$ & $0.126\pm 0.005\pm 0.002$ & $0.126\pm 0.005\pm 0.002$ &
$0.125\pm 0.005\pm 0.002$
\\ \hline
$\chi^2/\mbox{\rm d.o.f.}$ & $7.6/9$ & $7.7/11$ & $15/12$
\\ \hline\hline
$\sin^2{\theta^{\mbox{\rms lept}}_{\mbox{\rms eff}}}$ &
$0.2322\pm 0.0004{\,}^{+0.0001}_{-0.0002}$ &
$0.2323\pm 0.0003{\,}^{+0.0001}_{-0.0002}$ &
$0.2320\pm 0.0003{\,}^{+0.}_{-0.0002}$
\\
$1-M_W^2/M_Z^2$ & $0.2249\pm 0.0013{\,}^{+0.0003}_{-0.0002}$ &
$0.2250\pm 0.0013{\,}^{+0.0003}_{-0.0002}$ &
$0.2242\pm 0.0012{\,}^{+0.0003}_{-0.0002}$
\\
$M_W$ \,\, (GeV) & $80.28\pm 0.07{\,}^{+0.01}_{-0.02}$ &
$80.27\pm 0.06{\,}^{+0.01}_{-0.01}$ &
$80.32\pm 0.06{\,}^{+0.01}_{-0.01}$
\\ \hline
\end{tabular}
\end{center}
\end{table}

\setcounter{equation}{0}
\section{Quark Mixing}
\label{sec:cc-quarks}

Our knowledge of the charged-current parameters is unfortunately not so
good as in the neutral-current case. In order to measure the CKM matrix
elements, one needs to study semileptonic weak decays
$H\to H' l \bar\nu_l$, associated with the corresponding quark transition
$d_j\to u_i l^-\bar\nu_l$.
Since quarks are confined within hadrons, the decay amplitude
\bel{eq:T_decay}
T[H\to H' l \bar\nu_l]\,\approx\, {G_F\over\sqrt{2}} \,\bV_{\!\! ij}\,
\,\langle H'| \bar u_i \gamma^\mu (1-\gamma_5) d_j | H\rangle \,\,
\bar l \gamma_\mu (1-\gamma_5) \nu_l
\ee
always involves an hadronic matrix element of the weak left current.
The evaluation of this matrix element is a non-perturbative QCD problem
and, therefore, introduces unavoidable theoretical uncertainties.

Usually, one looks for a semileptonic transition
where the matrix element can be fixed at some kinematical point, by
a symmetry principle. This has the virtue of reducing the theoretical
uncertainties to the level of symmetry-breaking corrections and
kinematical extrapolations.
The standard example is a $0^-\to 0^-$ decay such as $K\to\pi l\nu$,
$D\to K l\nu$ or $B\to D l\nu$.
Only the vector current can contribute in this case:
\bel{eq:vector_me}
\langle P'(k')| \bar u_i \gamma^\mu d_j | P(k)\rangle \, = \, C_{PP'}\,
\left\{ (k+k')^\mu f_+(t) + (k-k')^\mu f_-(t)\right\} .
\ee
Here, $C_{PP'}$ is a Clebsh-Gordan factor and $t=(k-k')^2$.
The unknown strong dynamics is fully contained in the form factors
$f_\pm(t)$.
In the massless quark limit, the divergence of
the vector current is zero; thus, $f_-(t)=0$ and, moreover, $f_+(0)=1$
because the associated flavour charge is a conserved quantity.
Therefore, one only needs to estimate the corrections induced by the
finite values of the quark masses.

Since
$(k-k')^\mu\,\bar l\gamma_\mu (1-\gamma_5)\nu_l\sim m_l$, the contribution
of $f_-(t)$ is kinematically suppressed in the $e$ and $\mu$ modes.
The decay width can then be written as
\bel{eq:decay_width}
\Gamma(P\to P' l \nu)
\, =\, {G_F^2 M_P^5\over 192\pi^3}\, |\bV_{\!\! ij}|^2\, C_{PP'}^2\,
|f_+(0)|^2\, \cI\, \left(1+\delta_{\mbox{\rms RC}}\right) ,
\ee
where $\delta_{\mbox{\rms RC}}$ is an electroweak radiative
correction factor and $\cI$ denotes a phase-space integral,
which in the $m_l=0$ limit takes
the form
\bel{eq:ps_integral}
\cI\,\approx\,\int_0^{(M_P-M_{P'})^2} {dt\over M_P^8}\,
\lambda^{3/2}(t,M_P^2,M_{P'}^2)\,
\left| {f_+(t)\over f_+(0)}\right|^2 .
\ee

The usual procedure to determine $|\bV_{\!\! ij}|$ involves three steps:
\begin{enumerate}
\item Measure the shape of the $t$ distribution. This fixes the ratio
$|f_+(t)/f_+(0)|$ and therefore determines $\cI$.
\item Measure the total decay width $\Gamma$. Since $G_F$ is already known
from $\mu$ decay, one gets then an
experimental value for the product $|f_+(0)|\, |\bV_{\!\! ij}|$.
\item Get a theoretical prediction for $f_+(0)$.
\end{enumerate}
The important point to realize is that theoretical input is
always needed. Thus, the accuracy of the $|\bV_{\!\! ij}|$
determination is limited by our ability to  calculate the relevant
hadronic input.

\subsection{V$_{\! ud}$}
\label{subsec:Vud}

The most accurate measurement of $\bV_{\!\! ud}$ is done with
superallowed nuclear $\beta$ decays of the Fermi type [$0^+\to 0^+$],
where the nuclear matrix element $\langle N'|\bar u\gamma^\mu d|N\rangle$
can be fixed by vector-current conservation.
The CKM factor is obtained through the relation \cite{MA:91},
\bel{eq:ft_value}
|\bV_{\!\! ud}|^2\, =\,
{\pi^3\ln{2}\over ft\, G_F^2 m_e^5\, (1+\delta_{\mbox{\rms RC}})}
\, = \, {(2984.4\pm 0.1)\,\mbox{\rm s}\over ft\,(1+\delta_{\mbox{\rms RC}})} ,
\ee
where the factor $ft$ denotes a ``comparative half-life'' corrected
for phase-space and Coulomb effects \cite{HA:90}.
In order to obtain $|\bV_{\!\! ud}|$, one needs to perform a careful analysis
of radiative corrections \cite{MS:86},
including both short-distance contributions
$\Delta_{\mbox{\rms inner}}=0.0234\pm 0.0012$, and
nucleus-dependent corrections $\Delta_{\mbox{\rms outer}}\equiv
\delta_{\mbox{\rms RC}}-\Delta_{\mbox{\rms inner}}$.
These radiative corrections are quite large,
$\delta_{\mbox{\rms RC}}\sim 3$-4\%, and have a crucial role in order
to bring the results from different nuclei into good agreement.
The final result quoted by the Particle Data Group \cite{pdg:94} is
\bel{eq:Vud}
|\bV_{\!\! ud}|\, =\, 0.9744\pm 0.0010 \, .
\ee

An independent determination can be obtained from neutron decay,
$n\to p e^-\nu_e$.
The axial current also contributes in this case; therefore, one needs
the experimental determination of the axial coupling
$g_A= 1.2573\pm 0.0028$ \cite{pdg:94}.
The measured neutron lifetime,
$\tau_n = 887.0\pm 2.0$~s, implies \cite{MA:91}:
\bel{eq:n_decay}
|\bV_{\!\! ud}|\, =\,
\left\{ {(4904.0\pm 5.0)\,\mbox{\rm s}\over \tau_n (1+3 g_A^2)}\right\}^{1/2}
\, =\, 0.981\pm 0.002 \, ,
\ee
which is bigger than
\eqn{eq:Vud}.
Thus, a better measurement of $g_A$ and $\tau_n$ is needed.

The pion $\beta$-decay $\pi^+\to\pi^0 e^+\nu_e$ offers a
cleaner way to measure
$|\bV_{\!\! ud}|$. It is a pure vector transition, with very small theoretical
uncertainties. Unfortunately, owing to the kinematical suppression,
it has a small branching fraction.
The present experimental value is not very precise,
Br=$(1.025\pm 0.034)\times 10^{-8}$; it implies
$|\bV_{\!\! ud}| = 0.968\pm 0.018$.
An accurate measurement of this transition would be very valuable.

\subsection{V$_{\! us}$}
\label{subsec:Vus}

The decays $K^+\to\pi^0 l^+\nu_l$ and $K^0\to\pi^-l^+\nu_l$
are ideal for measuring $|\bV_{\!\! us}|$, because the relevant hadronic
form factors are well understood.
$SU(3)$ breaking corrections are very suppressed \cite{AG:64}
and isospin violations
can be easily taken into account:
$f_+^{K^0\pi^-}(0)=1 + \cO[(m_s-m_u)^2]$;
$f_+^{K^+\pi^0}(0)=1 + 3 (m_d-m_u)/(4m_s-2m_u-2m_d)+\ldots$
Moreover, higher-order corrections have been estimated
\cite{GL:85}, using Chiral Perturbation Theory methods \cite{GL:85,PI:93a}.
The resulting  values \cite{GL:85},
$f_+^{K^0\pi^-}(0)=0.977$ and
$f_+^{K^+\pi^0}(0)/f_+^{K^0\pi^-}(0)=1.022$, should be compared with
the experimental ratio \cite{pdg:94}
$|f_+^{K^+\pi^0}(0)/f_+^{K^0\pi^-}(0)|=1.028\pm 0.010$.
The accurate calculation of these quantities allows to extract
\cite{LR:84} a precise determination of $|\bV_{\!\! us}|$:
\bel{eq:Vus}
|\bV_{\!\! us}|\, =\, 0.2196\pm 0.0023 \, .
\ee

The analysis of semileptonic hyperon decay data can also
provide information on $|\bV_{\!\! us}|$. However,  the theoretical
uncertainties are much larger, owing to the first-order
$SU(3)$-breaking effects in the axial-vector couplings \cite{DHK:87}.
The Particle Data Group \cite{pdg:94} quotes the result
$|\bV_{\!\! us}|\, =\, 0.222\pm 0.003$.
The average with \eqn{eq:Vus} gives the final value:
\bel{eq:V_us}
|\bV_{\!\! us}|\, =\, 0.2205\pm 0.0018 \, .
\ee

\subsection{V$_{\! cd}$ and V$_{\! cs}$}
\label{subsec:Vcd}

The value of $|\bV_{\!\! cd}|$ is deduced
from deep inelastic $\nu_\mu$ and $\bar\nu_\mu$ scattering data,
by measuring the dimuon production rates off valence $d$ quarks;
i.e. $\nu_\mu d\to\mu^- c$ with the charm quark detected
through $c\to \mu^+\nu_\mu d$ or $\mu^+\nu_\mu s$.
One gets in this way, the product \cite{pdg:94}
$\overline{B}_c\,|\bV_{\!\! cd}|^2 = (0.47\pm 0.05)\times 10^{-2}$,
where $\overline{B}_c$ is the average semileptonic branching fraction
of the produced charmed hadrons.
Using $\overline{B}_c=0.113\pm 0.015$ \cite{pdg:94}, yields
\bel{eq:Vcd}
|\bV_{\!\! cd}|\, =\, 0.204\pm 0.017 \, .
\ee
%


Similarly, one could extract $|\bV_{\!\! cs}|$ from
$\nu_\mu s\to\mu^- c$ data. The resulting values depend, however,
on assumptions about the strange quark density in the parton-sea.
Making the conservative assumption that the strange quark-sea does not
exceed the value corresponding to an $SU(3)$ symmetric sea, leads
to the lower bound  \cite{AB:82} $|\bV_{\!\! cs}|>0.59$.

Better information is obtained from the decay $D\to\bar K e^+\nu_e$.
The measured $t$ distribution \cite{pdg:94}
can be fitted with the parametrization
$f_+^D(t)/f_+^D(0) = M^2/(M^2-t)$ and $M=2.1$ GeV; this determines
the corresponding integral $\cI$.
Using $\Gamma(D\to\bar K e^+\nu_e)=(0.762\pm 0.055)\times 10^{11}$
s$^{-1}$, one gets \cite{pdg:94}:
\bel{eq:Vcs}
|f_+^D(0)|\, |\bV_{\!\! cs}|\, =\, 0.704\pm 0.026 \, .
\ee
The status of our theoretical understanding of charm form factors
is quite crude. The symmetry arguments are not very helpful here,
because the charm-quark mass is too heavy for using the $SU(4)$
massless limit, and, at the same time, is too light to believe the
naive results obtained in the limit $m_c\to\infty$.
Symmetry-breaking corrections are very important.
The conservative assumption $|f_+^D(0)|<1$, implies
$|\bV_{\!\! cs}|>0.62$.
The Particle Data Group \cite{pdg:94} takes the range
$|f_+^D(0)|= 0.7\pm 0.1$, which covers the main part of the existing
calculations, and quotes:
\bel{eq:V_cs}
|\bV_{\!\! cs}|\, =\, 1.01\pm 0.18 \, .
\ee

Theoretical uncertainties are largely avoided by taking ratios,
such as $\Gamma(D\to\pi l\nu_l)/\Gamma(D\to \bar K l\nu_l)$,
where the form-factor uncertainty is reduced to the level of
$SU(3)$ breaking. The present measurements of the $D\to\pi l\nu_l$
decays are still too poor to be competitive. However, a 1\% measurement
of these semileptonic ratios seems possible at a future
tau-charm factory; this would allow a precise determination of
$|\bV_{\!\! cd}|/|\bV_{\!\! cs}|$.

\subsection{V$_{\! cb}$ and V$_{\! ub}$}
\label{subsec:Vcb}

Assuming that the inclusive semileptonic decay width of a bottom hadron
is given by the corresponding quark decay $b\to c l\nu_l$, the
magnitude of $|\bV_{\!\! cb}|$ can be determined from the ratio of
the measured semileptonic branching ratio and lifetime.
However, since $\Gamma(b\to c l\nu_l)\propto m_b^5$, this method is
very sensitive to the not so well-known value of the $b$-quark mass.
Moreover, higher-order QCD corrections are sizeable.
A recent experimental summary \cite{PA:94} quotes the results:
$|\bV_{\!\! cb}| = 0.039\pm 0.001 \pm 0.005$ from $\Upsilon(4S)$ data,
and
$|\bV_{\!\! cb}|= 0.042\pm 0.002\pm 0.005$ from LEP data,
where the second error gives an educated guess of the theoretical uncertainty.

The cleanest determination of $|\bV_{\!\! cb}|$ uses the decay
$B\to D^* l \bar\nu_l$ \cite{NE:91},
where the relevant hadronic form factor
[$\cF(v_B\cdot v_{D^*})$]
can be controlled at the level of a few per cent, close
to the zero-recoil region.
In the infinite $B$-mass limit, the normalization of this form factor
at zero recoil is fixed to be one, and the leading $1/M$ corrections
vanish \cite{LU:90}
due to heavy-quark symmetry; thus the theoretical uncertainty
is of order $1/M^2$ and therefore in principle small.
The calculated short-distance QCD corrections and the present estimates
of the $1/M^2$ contributions result in
$\cF(1) = 0.93\pm 0.03$ \cite{NE:94}, implying
\bel{eq:V_cb}
|\bV_{\!\! cb}|= 0.040\pm 0.003 \, .
\ee
%


The present determination of $|\bV_{\!\! ub}|$ is based on measurements
of the lepton momentum spectrum in inclusive $B\to X_q l\bar\nu_l$
decays, where $X_q$ is any hadronic state containing a quark $q=c$ or $u$.
The method is very sensitive to the assumed theoretical spectrum near the
kinematic limit for $B\to D l\bar\nu_l$.
Using different models to estimate the systematic theoretical
uncertainties, the analyses of the experimental data give
\cite{pdg:94}
\bel{eq:Vub}
\left| {\bV_{\!\! ub} / \bV_{\!\! cb}}\right|
\, =\,  0.08\pm 0.02 \, .
\ee
Eq.~\eqn{eq:V_cb}, implies then
$|\bV_{\!\! ub}| = 0.003\pm 0.001$.

\subsection{Unitarity}
\label{subsec:unitarity}

The present status of direct $\bV_{\!\! ij}$ determinations can
be easily summarized:
\bi
\item The light-quark mixings $|\bV_{\!\! ud}|$ and $|\bV_{\!\! us}|$
are rather well-known (0.1\% and 0.8\% accuracy, respectively).
Moreover, since the theory is good, improved values could be obtained
with better data on semileptonic $\pi^+$ and $K$ decays.
\item $|\bV_{\!\! cd}|$ and $|\bV_{\!\! cs}|$  are very badly known
(8\% and 18\% accuracy, respectively). This could be largely improved
at a tau-charm factory. Much better theoretical input is needed.
\item $|\bV_{\!\! cb}|$ and $|\bV_{\!\! ub}|$ are also badly known
(8\% and 33\% accuracy, respectively). However, there are good
theoretical tools available. Thus, better determinations could be
easily performed at a $B$ factory.
\item Nothing is known about the CKM mixings involving the top quark.
\ei

The entries of the first row are already accurate enough to perform
a sensible test of the unitarity of the CKM matrix:
\bel{eq:unitarity_test}
|\bV_{\!\! ud}|^2 + |\bV_{\!\! us}|^2 + |\bV_{\!\! ub}|^2 \, = \,
0.9981\pm 0.0027 \, .
\ee
It is important to notice that radiative corrections play here a
crucial role. If one uses $|\bV_{\!\! uj}|$ values determined without
radiative corrections, the result \eqn{eq:unitarity_test}
changes to $1.0384\pm 0.0027$, giving an apparent violation
of unitarity (by many $\sigma$'s) \cite{MA:91}.

Imposing the unitarity constraint $\bV \bV^\dagger = \bV^\dagger \bV =1$
(and assuming only three generations)
one can get a more precise picture of the CKM matrix.
The 90\% confidence limits on the magnitude of the CKM matrix elements are
then \cite{pdg:94}:
\bel{eq:CKM_values}
\bV \, = \, \left[
\begin{array}{ccc}
0.9747 \,\,\mbox{\rm to}\,\, 0.9759  & 0.218 \,\,\mbox{\rm to}\,\, 0.224 &
0.002 \,\,\mbox{\rm to}\,\, 0.005 \\
0.218 \,\,\mbox{\rm to}\,\, 0.224 & 0.9738 \,\,\mbox{\rm to}\,\, 0.9752 &
0.032 \,\,\mbox{\rm to}\,\, 0.048 \\
0.004 \,\,\mbox{\rm to}\,\, 0.015 & 0.030 \,\,\mbox{\rm to}\,\, 0.048 &
0.9988 \,\,\mbox{\rm to}\,\, 0.9995
\end{array}
\right] .
\ee
The ranges given here are slightly different from (but consistent with)
the direct determinations mentioned before.

The resulting CKM matrix shows a hierarchical pattern, with the
diagonal elements being very close to one, the ones connecting the
two first generations having a size
$\lambda\equiv |\bV_{\!\! us}|= 0.2205\pm 0.0018$,
the mixing between the second and third families being of order
$\lambda^2$, and the mixing between the first and third quark flavours
having a much smaller size of about $\lambda^3$.
It is then quite practical to use the
approximate parametrization \cite{WO:83}:
\be\label{eq:wolfenstein}
\bV\, =\,
\left[ \matrix{\displaystyle \ 1- {\lambda^2 \over 2}
\hfill&
\displaystyle \ \ \ \ \ \ \lambda \hfill&
\displaystyle \ \ \ \ \ A\lambda^
3(\rho  - i\eta) \hfill \cr\displaystyle
\hfill& \displaystyle \hfill&
\displaystyle \hfill \cr\displaystyle \ \ \ -\lambda
\hfill& \displaystyle \ \
\ \ \ 1 -{\lambda^ 2 \over 2} \hfill& \displaystyle
\ \ \ \ \ A\lambda^ 2 \hfill
\cr\displaystyle \hfill& \displaystyle \hfill&
\displaystyle \hfill
\cr\displaystyle \ A\lambda^ 3(1-\rho -i\eta)
\hfill& \displaystyle \ \ \ \ \
-A\lambda^ 2 \hfill& \displaystyle
\ \ \ \ \ \ \ \ \ 1 \hfill \cr} \right]\
+\ O\left(\lambda^ 4 \right) \, ,
\ee
where $A=|\bV_{\!\! cb}|/\lambda^2 = 0.82\pm 0.06$,
and
\bel{eq:circle}
\sqrt{\rho^2+\eta^2} \, = \,
\left|{\bV_{\!\! ub}\over \lambda \bV_{\!\! cb}}\right|
\, =\, 0.36\pm 0.09 \, .
\ee

\subsection{Indirect determinations}
\label{subsec:indirect}

Additional information can be obtained from
flavour-changing neutral-current transitions, occurring at the 1--loop
level. An important example is provided by
the mixing between the $B^0$ meson and its antiparticle.
This process occurs through the exchange of two $W$'s between the
fermionic lines, the so-called box diagrams.
The mixing amplitude is proportional to
\bel{eq:mixing}
\langle\bar B_d^0 | \cH|B_0\rangle\,\sim\,
\sum_{ij}\, \bV_{\!\! id}^{\phantom{*}}\bV_{\!\! ib}^*
\bV_{\!\! jd}^*\bV_{\!\! jb}^{\phantom{*}}\,
S(r_i,r_j) ,
\ee
where $S(r_i,r_j)$ is a loop function which depends on the masses
[$r_i\equiv m_i^2/M_W^2$]
of the up-type quarks running along the internal lines.
Owing to the unitarity of the CKM matrix, the mixing amplitude vanishes
for equal (up-type) quark masses (GIM mechanism); thus the effect
is proportional to the mass splittings between the $u$, $c$ and $t$ quarks.
Since the different CKM factors have all a similar size,
$\bV_{\!\! ud}^{\phantom{*}}\bV_{\!\! ub}^*\sim
\bV_{\!\! cd}^{\phantom{*}}\bV_{\!\! cb}^*\sim
\bV_{\!\! td}^{\phantom{*}}\bV_{\!\! tb}^*\sim A\lambda^3$,
the final amplitude
is completely dominated by the top contribution;
i.e. $\langle\bar B_d^0 | \cH|B_0\rangle\,\sim\, |\bV_{\!\! tb}|^2 S(r_t,r_t)$.

\begin{figure}[b]  
\centerline{\mbox{\epsfysize=4.0cm\epsffile{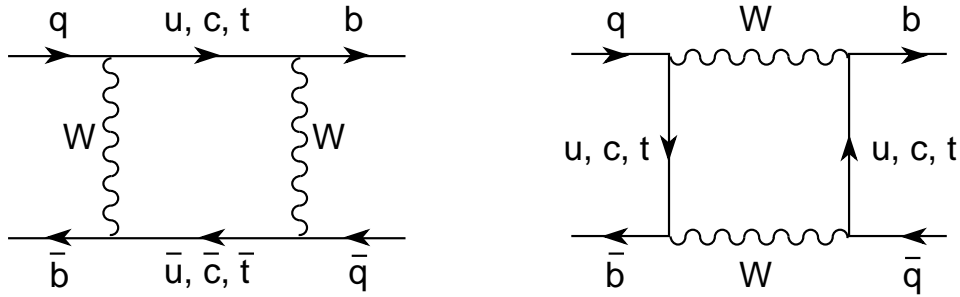}}}
\caption{$B^0$-$\bar B^0$ mixing diagrams.}
\label{fig:boxdia}
\end{figure}

One can then determine $|\bV_{\!\! td}|$ from the measured mixing
$x_d\equiv\Delta M_{B_d}/\Gamma_{B_d}=0.76\pm 0.06$ \cite{FO:94}.
Unfortunately,  one also needs to know the hadronic
matrix element of the 4-quark operator
$(\bar b_L\gamma^\mu d_L) (\bar b_L\gamma_\mu d_L)$ between the
$B^0$ and $\bar B^0$ states. This is again
a non-perturbative QCD problem,
which introduces a big theoretical uncertainty.
The most recent analysis \cite{PP:94} gets:
\bel{eq:V_td}
 |\bV_{\!\! td}|\, = \, 0.007{\,}^{+0.003}_{-0.002} \, ,
\ee
in good agreement with (but more precise than) the value obtained from the
unitarity constraint in Eq.~\eqn{eq:CKM_values}.
In terms of the $(\rho,\eta)$ parametrization of
Eq.~\eqn{eq:wolfenstein},
this gives
\bel{eq:circle_t}
\sqrt{(1-\rho)^2+\eta^2} \, = \,
\left|{\bV_{\!\! td}\over \lambda\bV_{\!\! cb}}\right|
\, = \, 0.8{\,}^{+0.3}_{-0.2}  \, .
\ee
Notice that together with \eqn{eq:circle}, this result implies
$\eta\not=0$; although errors are still too large to make any
strong statement. Thus, if CKM unitarity is assumed, it is possible
to establish the existence of CP violation from CP conserving measurements.
A more direct constraint on the parameter $\eta$ is provided \cite{PP:94} by
the measured CP violation in the $K^0$-$\bar K^0$ system, $\varepsilon_K$.

\setcounter{equation}{0}
\section{Charged-Current Lepton Universality}
\label{sec:cc-leptons}

In the SM, the $W$ couples with the same strength $g/(2\sqrt{2})$
to all charged fermionic currents (up to CKM mixing factors in the
quark sector).
The universality of the leptonic couplings can be easily tested, by
allowing these couplings to depend on the considered lepton flavour
and comparing several leptonic and semileptonic decays.

The ratio $R_{e/\mu}$ of the two semileptonic $\pi^-\to l^-\bar\nu_l$
decay modes is proportional to $|g_e/g_\mu|^2$. From Eq.~\eqn{eq:r_e_mu},
one inmediately gets:
\bel{eq:univ_e_mu}
\left| {g_\mu / g_e} \right| \, = \, 1.0021\pm 0.0016 \, .
\ee
A less accurate value,
$\left| g_\mu/ g_e \right| = 1.004\pm 0.009 $,
is obtained from the ratio
$B_{\tau\to\mu}/B_{\tau\to e} = 0.9800\pm 0.017$ \cite{pdg:94},
where $B_{\tau\to l}\equiv\Gamma(\tau^-\to l^-\bar\nu_l\nu_\tau)$.

Comparing
$\Gamma(\tau^-\to e^-\bar\nu_e\nu_\tau)$ and
$\Gamma(\mu^-\to e^-\bar\nu_e\nu_\mu)$ [see Eq.~\eqn{eq:mu_lifetime}],
one can test $g_\tau$:
\bel{eq:univ_mu_tau}
\left| {g_\tau\over g_\mu} \right| \, = \,
\left\{ B_{\tau\to e}\,
{\tau_\mu m_\mu^5\, f(m_e^2/m_\mu^2) \,
(1+\delta^\mu_{\mbox{\rms RC}})\over
\tau_\tau m_\tau^5 \, f(m_e^2/m_\tau^2) \,
(1+\delta^\tau_{\mbox{\rms RC}})}
\right\}^{1/2}
\, = \, 0.997\pm 0.007 \, .
\ee
Other (less precise) tests of lepton universality are obtained
from $p\bar p$ collider data on leptonic $W$ decays, and
from the ratio of
the
$\Gamma[\tau^-\to\nu_\tau\pi^- (K^-)]$ and
$\Gamma[\pi^-(K^-)\to l^-\bar\nu_l]$ decay widths \cite{PI:93}.

The $V-A$ structure of the $\tau$ charged current can also be studied,
following the same kind of analysis performed for $\mu$ decay
(see Sect.~\ref{subsec:mu-decay}).
Unfortunately, the data is still not accurate enough to determine the
interaction. Assuming that the $\tau\nu_\tau W$ vertex is a linear
combination of vector and axial currents,
$g_V^{(\tau)} V^\mu - g_A^{(\tau)} A^\mu$,
and using the SM $V-A$ form for the other $W$ couplings, one
gets the constraints \cite{PI:93,AR:94}:
\bel{eq:michel}
\left| {g_V^{(\tau)}- g_A^{(\tau)}\over g_V^{(\tau)}+ g_A^{(\tau)}}\right|
\, < \, 0.37 \qquad (95\% \mbox{\rm CL}), \qquad\qquad
{2 g_V^{(\tau)} g_A^{(\tau)}\over \left|g_V^{(\tau)}\right|^2+
\left|g_A^{(\tau)}\right|^2}
\, = \, 1.022\pm 0.041 .
\ee

\setcounter{equation}{0}
\section{Summary and Outlook}
\label{sec:summary}

The SM provides a beautiful theoretical framework which is able to
accommodate all our present knowledge on electroweak interactions.
It is able to explain any single experimental fact,  and in some cases,
such as the neutral-current sector, has succesfully passed very precise
tests at the 0.1\% to 1\% level.
However, there are still  pieces of the SM Lagrangian which so-far
have not been experimentally analyzed in any precise way.
Moreover, the SM leaves many unaswered questions and contains too many
free parameters to qualify as an ultimate fundamental theory.
Clearly, new physics should exist.

  The discovery of the top quark \cite{CDF:94} is awaiting confirmation.
In addition to complete the SM fermionic structure in the third
generation,
an accurate measurement of $m_t$ is needed to improve the significance of
present neutral-current analyses at the $Z$ peak.
Together with a much better meaurement of $\Gamma(Z\to\bar bb )$,
that would provide a non-trivial consistency test of the SM at the
quantum level, including effects related with the longitudinal gauge-boson
polarization.

The gauge self-couplings will be investigated at LEP II, through the
study of the $e^+e^-\to W^+W^-$ production cross-section.
The $V-A$ ($\nu_e$-exchange in the $t$ channel) contribution generates
an unphysical growing of the  cross-section with the centre-of-mass energy,
which is compensated through a delicate gauge cancellation with the
$e^+e^-\to\gamma, Z\to W^+W^-$ amplitudes.
This offers a good way to test the gauge-boson self-interactions.
The study of this process will also provide a more accurate measurement of
$M_W$, allowing to improve the precision of present LEP I analyses.

The Higgs particle is the main missing block of the SM framework.
The present experimental lower bound is \cite{pdg:94}
\bel{eq:Higgs_exp}
M_H \, > \, 58.4\,\,\mbox{\rm GeV}   \qquad\qquad (95\%\, \mbox{\rm CL}) .
\ee
LEP II first and later LHC will try to find out wether such scalar field
exists.
Note that the present succesful tests of the $\rho_0 = 1$ prediction only
provide a confirmation of the assumed pattern of SSB, but do not prove
the minimal Higgs mechanism embedded in the SM.

The Higgs width increases very fast with its mass
[$\Gamma(H\to W^+W^-, ZZ)\sim G_F M_H^3$].
At $M_H\sim 1$ TeV, $\Gamma_H\sim M_H$; thus a heavy Higgs would look
experimentally
like a very broad resonant structure rather than as a ``fundamental''
peak.
In fact, since the $|\phi|^4$ coupling grows with $M_H$
($h\sim M_H^2$),
at such large masses the Higgs interactions
are very strong. For $M_H\geq\sqrt{2} v\approx 348$ GeV, $h\geq 1$
and the SM enters into a non-perturbative strong-coupling regime.
A naive resummation of higher-order corrections to the
$|\phi|^4$ vertex generates an effective ``running'' coupling $h(s)$,
which grows with the energy scale and blows up (Landau pole)
at $\sqrt{s}\sim v \exp{\left\{-3 M_H^2/(4\pi^2 v^2)\right\}}$.
The perturbative predictions become completely meaningless
above $s\sim M_H\sim 815$ GeV.
A similar phenomenon happens in the scattering of longitudinal
gauge bosons, where the tree-level $W^+_LW^-_L\to W^+_LW^-_L$
amplitude violates the unitarity limit for $M_H\geq 713$ GeV;
higher-order contributions (which obviously would restore unitarity)
are then huge, indicating again a non-perturbative regime.
Thus, the experimental investigation of the SSB mechanism at higher-energy
machines could provide hints of completely new phenomena.

The family structure and the pattern of fermionic masses and mixings
constitute a ``terra incognita'', where we know nothing else than
the empirical determinations of the relevant parameters.
Flavour factories (such as kaon, tau-charm, $B$ or even a futuristic
top factory) are needed in order to make an accurate investigation
of the properties of the different fermionic flavours.
A precise (and overconstrained) measurement of the quark-mixing parameters
would allow to test the unitarity structure of the CKM matrix.

CP-violation offers an interesting window into possible new physics.
The tiny violation of the CP symmetry observed in the kaon system,
can be parametrized through the CKM phase. However, a fundamental
explanation of the origin of this phenomena is lacking.
In the SM, all CP-violating effects should be explained by a single
parameter $\delta_{13}$; moreover, any signal should disappear in the
limit where any two equal-charge quarks become degenerate in mass
(the CP phase could then be rotated away by a field redefinition).
Thus, the SM makes very precise predictions
for CP-violating observables, which should be tested in appropriate
experiments.

Finally, the possibility of non-zero neutrino  massess,
and the associated lepton-flavour violation phenomena, should be
investigated.
Moreover, a better knowledge of the $\tau$-neutrino properties
is required. The existence of the $\nu_\tau$ as a different neutrino
flavour can be inferred from the measured invisible $Z$ width;
however, so far, nobody has been able to detect a single $\nu_\tau$
interaction.

Clearly, we need more experiments in order to learn what kind of
physics exists beyond the present SM frontiers.
We have, fortunately, a very promising and exciting future
ahead of us.

\section*{Acknowledgements}

My first contact with the SM was through the excellent lecture notes by
J.~Bernab\'eu and P.~Pascual \cite{BP:81}; its influence on some sections
of the present lectures is obvious.
The help of M.~Martinez, member of the LEP Electroweak Working Group,
has been very valuable in order to incorporate the results of
Ref.~\cite{lep:94}.
Finally, I would like to thank J.~Bernab\'eu and F.J.~Botella for their
useful comments on the manuscript.
This work has been supported in part by CICYT (Spain) under grant
No. AEN-93-0234.


\begin{Thebibliography}{99}

\bibitem{GL:61} S.L. Glashow, Nucl. Phys. 22 (1961) 579.

\bibitem{WE:67} S. Weinberg, Phys. Rev. Lett. 19 (1967) 1264.

\bibitem{SA:69} A. Salam, in {\em Elementary Particle Theory},
ed. N. Svartholm (Almquist and Wiksells, Stockholm, 1969), p. 367.

\bibitem{GIM:70} S.L. Glashow, J. Iliopoulos and L. Maiani,
Phys. Rev. D2 (1970) 1285.

\bibitem{sorrento}
  D. Espriu, {\em Perturbative QCD}, these proceedings; \\
  A. Pich, {\em Quantum ChromoDynamics}, lectures given at the
  1994 CERN-JINR Summer School (Sorrento, Italy, September 1994).

\bibitem{fgj:86}
   W. Fetscher, H.-J. Gerber and K.F. Johnson,
   Phys. Lett. B173 (1986) 102.

\bibitem{KS:59}
T. Kinoshita and A. Sirlin, Phys. Rev. 113 (1959) 1652.

\bibitem{pdg:94}
   Particle Data Group, {\em Review of Particle Properties},
   Phys. Rev. D50 (1994) 1173.

\bibitem{MS:93}
W.J. Marciano and A. Sirlin, Phys. Rev. Lett. 71 (1993) 3629.

\bibitem{DA:62}
G. Danby et al., Phys. Rev. Lett. 9 (1962) 36.

\bibitem{cabibbo}
   N. Cabibbo, Phys. Rev. Lett. 10 (1963) 531.

\bibitem{LY:60}
T.D. Lee and C.N. Yang, Phys. Rev. Lett. 4 (1960) 307.

\bibitem{HA:73} H.J. Hasert et al., 
Phys. Lett. 46B (1973) 121.

\bibitem{KI:90} T. Kinoshita (editor), {\em Quantum Electrodynamics},
Advanced Series on Directions in High Energy Physics, Vol. 7
(World Scientific, Singapore, 1990);\\
T. Kinoshita, Phys. Rev. D47 (1993) 5013.


\bibitem{goldstone} J. Goldstone, Nuov. Cim. 19 (1961) 154.

\bibitem{HI:66} P.W. Higgs, Phys. Rev. 145 (1966) 1156; \\
  T.W.B Kibble, Phys. Rev. 155 (1967) 1554.

\bibitem{TH:71} G. 't Hooft, Nucl. Phys. B33 (1971) 173.

\bibitem{cern:83}
G. Arnison et al. (UA1), Phys. Lett. B122 (1983) 103; B126 (1993) 398; \\
M. Banner et al. (UA2), Phys. Lett. B122 (1983) 476; B129 (1983) 130.

\bibitem{lep:94}
The LEP Collaborations and the LEP Electroweak Working Group,
CERN/PPE/94-187.

\bibitem{MW:94}
M. Demarteau et al., 
CDF/PHYS/CDF/PUBLIC/2552 and D0 NOTE 2115.

\bibitem{charm}
  J.J. Aubert et al., Phys. Rev. Lett. 33 (1974) 1404;\\
  J.E. Augustin et al., Phys. Rev. Lett. 33 (1974) 1406; 1453.

\bibitem{KM:73} M. Kobayashi and T. Maskawa, Prog. Theor. Phys. 42
  (1973) 652.

\bibitem{CDF:94}
  F. Abe et al (CDF), Phys. Rev. D50 (1994) 2966;
  Phys. Rev. Lett. 73 (1994) 225.

\bibitem{slc:94}
K. Abe et al. (SLD), Phys. Rev. Lett. 73 (1994) 25.

\bibitem{BU:89} H. Burkhardt et al., Z. Phys. C43 (1989) 497.

\bibitem{VE:77} M. Veltman, Nucl. Phys. B123 (1977) 89.

\bibitem{BPS:88} J. Bernab\'eu, A. Pich and A. Santamar\'{\i}a,
Phys. Lett. B200 (1988) 569; Nucl. Phys. B363 (1991) 326.

\bibitem{SI:80} A. Sirlin, Phys. Rev. D22 (1980) 971.

\bibitem{AC:75}
T. Appelquist and J. Carazzone, Phys. Rev. D11 (1975) 2856.

\bibitem{ABR:86} A.A. Akhundov, D. Yu. Bardin and T. Riemann,
Nucl. Phys. B276 (1986) 1; \\
W. Beenakker and W. Hollik, Z. Phys. C40 (1988) 141; \\
B.W. Lynn and R.G. Stuart, Phys. Lett. B252 (1990) 676.

\bibitem{BCS:93} J.M. Benlloch et al., Z. Phys. C59 (1993) 471.

\bibitem{charmII:94}
P. Vilain et al. (CHARM II), Phys. Lett. B335 (1994) 246.

\bibitem{QCD:94}
  A. Pich, {\em QCD Predictions for the $\tau$ Hadronic Width:
  Determination of $\alpha_s(M_\tau^2)$},
  to appear in Proc. QCD 94 Workshop (Montpellier, 1994)
  [Valencia preprint FTUV/94-71].

\bibitem{MA:91}
W.J. Marciano, Annu. Rev. Nucl. Part. Sci. 41 (1991) 469.

\bibitem{HA:90}
J.C. Hardy et al., Nucl. Phys. A509 (1990) 429.

\bibitem{MS:86}
W.J. Marciano and A. Sirlin, Phys. Rev. Lett. 56 (1986) 22;\\
A. Sirlin and R. Zucchini, Phys. Rev. Lett. 57 (1986) 1994;\\
A. Sirlin, Phys. Rev. D35 (1987) 3423; \\
W. Jaus and G. Rasche, Phys. Rev. D35 (1987) 3420; D41 (1990) 166.

\bibitem{AG:64} M. Ademollo and R. Gatto, Phys. Rev. Lett. 13 (1964) 264.

\bibitem{GL:85}
J. Gasser and H. Leutwyler, Nucl. Phys. B250 (1985) 465; 517; 539.

\bibitem{PI:93a} A. Pich, {\em Introduction to Chiral Perturbation Theory},
Proc. Fifth Mexican School of Particles and Fields (Guanajuato, Mexico, 1992),
eds. J.L. Lucio M. and M. Vargas, AIP Conference Proceedings 317
(New York, 1994), p. 95.

\bibitem{LR:84} H. Leutwyler and M. Roos, Z. Phys. C25 (1984) 91.

\bibitem{DHK:87}
J.F. Donoghue, B.R. Holstein and S.W. Klimt, Phys. Rev. D35 (1987) 934.

\bibitem{AB:82} H. Abramowicz et al., Z. Phys. C15 (1982) 19.

\bibitem{PA:94} R. Patterson, {\em Weak and rare decays},
to appear in Proc. of the 27th
 Int. Conf. on High Energy Physics (Glasgow, 1994).

\bibitem{NE:91} M. Neubert, Phys. Lett. B264 (1991) 455.

\bibitem{LU:90} M. Luke, Phys. Lett. B252 (1990) 447.

\bibitem{NE:94} M. Neubert, Phys. Lett. B338 (1994) 84.

\bibitem{WO:83} L. Wolfenstein, Phys. Rev. Lett. 51 (1983) 1945.

\bibitem{FO:94} R. Forty, {\em CP Violation and
$B^0$-$\bar B^0$ Mixing}, to appear in Proc. of the 27th
 Int. Conf. on High Energy Physics (Glasgow, 1994).

\bibitem{PP:94} A. Pich and J. Prades, Valencia preprint
FTUV/94-37.

\bibitem{PI:93} A. Pich, Nucl. Phys. B (Proc. Suppl.) 31 (1993) 213.

\bibitem{AR:94}
H. Albrecht et al. (ARGUS), Phys. Lett. B337 (1994) 383.

\bibitem{BP:81} J. Bernab\'eu and P. Pascual, {\em Electro-Weak Theory},
GIFT report (University Aut\'onoma of Barcelona, Bellaterra, 1981).

\end{Thebibliography}

\end{document}